\documentclass[letterpaper,twocolumn,10pt,
review,
anonymous,
]{article}
\usepackage{usenix2020-09}
\usepackage{amsfonts}
\usepackage{bbm}
\usepackage{booktabs}
\usepackage{makecell}
\usepackage{tikz}
\usepackage{adjustbox}
\usepackage{enumitem}
\usepackage{wrapfig}
\usepackage{subcaption}
\usepackage{xspace}
\usepackage[svgnames]{xcolor}

\usepackage[switch]{lineno}

\usepackage{algorithm}
\usepackage[noend]{algpseudocode}
\usepackage{mathrsfs}
\usepackage{amsmath}
\usepackage{amssymb}
\usepackage[makeroom]{cancel}
\usepackage{amsthm}

\usepackage{booktabs}

\usepackage{colortbl}

\usepackage{array}      
\usepackage{booktabs}   
\usepackage{breqn}
\usepackage{tabularx}

\newtheorem{thm}{Theorem}
\newtheorem{lem}[thm]{Lemma}

\usepackage{mdframed}
\newmdenv[
    linecolor=gray,
    linewidth=1pt,
    topline=false,
    bottomline=false,
    rightline=false,
    skipabove=2pt,
    skipbelow=1pt,
    leftmargin=0,
    rightmargin=10pt,
    innerleftmargin=3pt,
    innerrightmargin=0pt,
    innertopmargin=1pt,
    innerbottommargin=2pt,
    backgroundcolor=white
]{txtframe}
\newmdenv[
    linecolor=gray,
    linewidth=1pt,
    topline=true,
    bottomline=true,
    leftline=false,
    rightline=false,
    skipabove=1pt,
    skipbelow=1pt,
    leftmargin=10pt,
    rightmargin=10pt,
    innerleftmargin=2pt,
    innerrightmargin=2pt,
    innertopmargin=2pt,
    innerbottommargin=2pt,
    backgroundcolor=white
]{insightframe}

\usepackage[utf8]{inputenc}
\usepackage[T1]{fontenc}

\newif\ifshowcomment
\showcommenttrue

\newcommand{\sys}{{\textsc{NetNomos}}\xspace}

\newcommand{\tocite}[1]{[{\textcolor{red}{\textbf{!}}}]}
\newcommand{\toref}[1]{\textcolor{red}{\textbf{!}}}

\newcommand{\ie}{\emph{i.e.,} }
\newcommand{\eg}{\emph{e.g.,} }
\newcommand{\etc}{etc\@ifnextchar.{}{.\@}}
\usepackage[greek,english]{babel}

\ifshowcomment
\newcommand{\TODO}[1]{\textcolor{red}{{[\small\textsf{{TODO: #1}}}]}}
\newcommand{\NOTE}[1]{\textcolor{orange}{{[\small\textsf{{NOTE: #1}}}]}}
\else
\newcommand{\TODO}[1]{}
\newcommand{\NOTE}[1]{}
\fi
\newcommand{\x}{$\times$\xspace}

 \newcommand{\myitem}[1]{\vspace*{0.02in}\noindent\textbf{#1}}
\newcommand{\remove}[1]{}

\newcommand{\mypar}[1]{{\noindent\bf #1.\ }}

\newcommand{\assign}[0]{\mathrel{\mathop:}=}

 \newcommand{\circlewhite}[1]{%
 \begin{tikzpicture}[baseline=(char.base)]
   \node[draw,circle,inner sep=0.5pt, fill=white, text=black] (char){\small #1};
 \end{tikzpicture}%
 }

 \usepackage{tcolorbox}
\tcbuselibrary{listings}
\usepackage{graphicx}

\newtcolorbox{promptbox}{
  colback=gray!5,
  colframe=gray!60,
  listing only,
  listing options={
    basicstyle=\ttfamily\small,
    breaklines=true,
    columns=fullflexible
  }
}

\usepackage{tikz}
\usepackage{amsmath}

\usepackage{iftex}
\ifPDFTeX
  \input{glyphtounicode}
\fi

\usepackage[available,functional,reproduced]{usenixbadges}

\begin{document}

\date{}


\title{Making Logic a First-Class Citizen in Generative ML for Networking}

\author{
{\rm Hongyu H\`e}\\
Princeton University
\and
{\rm Minhao Jin}\\
Princeton University
\and
{\rm Maria Apostolaki}\\
Princeton University
} 

\maketitle
\pagestyle{empty}

\begin{abstract}

Generative ML models are increasingly popular in networking for tasks such as telemetry imputation, prediction, and synthetic trace generation. Despite their capabilities, they suffer from two shortcomings: \emph{(i)} their output is often visibly violating well-known networking rules, which undermines their trustworthiness; and \emph{(ii)} they are difficult to control, frequently requiring retraining even for minor changes.

To address these limitations and unlock the benefits of generative models for networking, we propose a new paradigm for integrating explicit network knowledge, in the form of first-order logic rules, into ML models used for networking tasks. Rules capture well-known relationships among observed signals, \eg that increased latency precedes packet loss.
While the idea is conceptually straightforward, its realization is challenging: networking knowledge is rarely formalized into rules, and naively injecting rules into ML models often hampers their effectiveness. This paper introduces \sys, a multi-stage framework that \emph{(i)} learns rules directly from data (\eg measurements); \emph{(ii)} filters them to select semantically meaningful ones; and \emph{(iii)} enforces them through collaborative generation between an ML model and a Satisfiability Modulo Theories (SMT) solver.

We show that \sys learns diverse, meaningful rules from four real-world datasets and is 1.6--6.5$\times$ more scalable than DuoAI, a state-of-the-art (SOTA) rule-learning method.
By enforcing these rules on a generic GPT-2 model, \sys achieves performance on par with or even surpassing specialized SOTA systems such as Zoom2Net and NetShare across three networking tasks: telemetry imputation, traffic forecasting, and synthetic data generation.

\end{abstract}


\section{Introduction}

Generative machine learning (ML) models are increasingly popular for networking tasks such as measurement imputation, prediction, and synthetic trace generation. Their appeal lies in their adaptability, their capacity to learn directly from abundant network data, and their efficiency during inference.

Despite their potential, generative models often fall short in networking applications. They lack the strict correctness guarantees essential for network operations and offer limited controllability, severely restricting their real-world utility. Consequently, they can make egregious errors that degrade downstream performance, erode user trust, and are notoriously difficult to fix. For example, even state-of-the-art (SOTA) synthetic generators might produce UDP packets with TCP flags, skewing network sketch evaluations, and frustrating operators. Worse, fixing these mistakes typically requires expensive fine-tuning or complete retraining, assuming a fix is even possible~\cite{sivaroopan2025comprehensive, jiang2024netdiffusion, guthula2023netfound}.

Infusing generative models with domain knowledge can steer them away from nonsensical outputs, guarantee compliance with established networking principles, and reduce training overhead by eliminating the need to learn basic rules purely from data. However, this infusion is exceptionally challenging. First, networking knowledge is rarely formalized in a machine-readable format. While written sources like RFCs could theoretically be converted into such a format~\cite{Basin2025Gap}, they typically only describe a single protocol in isolation, ignoring the complex cross-layer interactions of real-world networks. Second, enforcing rules within ML pipelines is non-trivial. Existing constraint methods are often either too rigid—stifling the model's predictive ability (a known issue in large language models~\cite{zhou2024rulearena,tam2024let})—or too loose, failing to provide meaningful guarantees. For example, best-effort enforcement in systems like NetDiffusion~\cite{jiang2024netdiffusion} still permits frequent rule violations, as we demonstrate in experiments (\S\ref{sec:eval}).

This paper introduces \sys\footnote{Nomos comes from a Greek word meaning rule or law, reflecting \sys’s goal of learning and enforcing the underlying network principles.}, a novel, modular pipeline for designing controllable network data generators that provide strict correctness guarantees. \sys bridges this gap by combining the predictive power of ML with the underlying formal rules governing network data.

\myitem{\sys's Knowledge Mining.} Since formalized networking knowledge is scarce, we extract rules directly from data (\eg packet headers, measurements), which naturally obey constraints dictated by protocols and deployment policies. Viewing data samples as feasible solutions to these hidden constraints, \sys automatically mines the underlying rules. To balance scalability and expressiveness, \sys defines a first-order logic (FOL) grammar over observable variables and reduces the rule-learning task to the minimum hitting set problem~\cite{hartmanis1982computers}.

\myitem{\sys's Filtering.} A critical challenge is that rules learned purely from data can be coincidental and lack true semantic meaning. However, unlike opaque ML weights, logic rules are discrete and auditable. \sys uses LLMs—grounded in networking texts—to filter and select meaningful rules. By restricting the LLM to choose only from mathematically mined rules, we ensure that consistency guarantees are preserved.

\myitem{\sys's Knowledge Enforcement.} \sys introduces a new approach to generation by embedding a Satisfiability Modulo Theories (SMT) solver directly into the language model's token-by-token generation process. Unlike prior methods that bake rules into training or attempt post-hoc corrections, \sys provides strict guarantees with minimal invasion to the underlying model, preserving its fidelity. This inference-time enforcement also allows models to be easily repurposed for new tasks by simply swapping the rule set, eliminating the need for retraining.

We evaluate \sys's ability to scalably learn meaningful rules and improve data generation across diverse datasets. To support this, we built and open-sourced new benchmarks for networking rules~\cite{netnomos} alongside an extensive suite of baselines from three domains. Our results show that traditional database and ML approaches (\eg FastDC~\cite{chu2013fastdc,pena2022fastdc++}, H-Mine~\cite{pei2007hmine}, FlowChronicle~\cite{ferraiuolo2018hyperflow}) lack the expressiveness needed to formalize network knowledge. Furthermore, \sys is $1.6\times$ to $6.5\times$ more scalable than DuoAI~\cite{yao2022duoai}, a SOTA, equally expressive rule-learning approach from the distributed systems domain.

Finally, we demonstrate the full \sys pipeline using a lightweight language model (GPT-2) across three use cases: telemetry imputation, synthetic data generation, and prediction. \sys reliably produces compliant network data—unlike heavily tailored SOTA systems (Zoom2Net~\cite{gong2024zoom2net}, NetDiffusion~\cite{jiang2024netdiffusion}, NetShare~\cite{yin2022netshare})—while matching or exceeding their performance metrics. Crucially, enforcing rules exclusively at inference is sufficient to tailor a single, basic GPT-2 model to all three tasks. This highlights the power of decoupling network knowledge extraction from ML inference, rather than relying on an opaque, end-to-end black box.

\section{Motivation}

We first examine two networking use cases that demand a neurosymbolic approach. While generative models have unlocked new capabilities for these use cases, they still require explicit knowledge infusion to be truly reliable. We then outline the fundamental pitfalls of prior designs that fail to effectively bridge machine learning and symbolic reasoning, hence fail to realize a truly neurosymbolic approach.

\subsection{Motivating Use Cases}\label{sec:usecase}

\myitem{Synthetic data generation.} Generative models can produce synthetic network traces (\eg packet headers and timestamps) by learning the latent structures and dependencies in real traffic~\cite{lin2019generating,yin2022netshare,jiang2024netdiffusion,chu2025netssm,jin2025tracebleed}. These synthetic data generators (SynGens) allow third parties to optimize downstream applications without accessing sensitive raw data. However, most SynGens rely on end-to-end pipelines that force the model to learn all networking rules purely from data. This approach is data-hungry, computationally expensive, and offers no correctness guarantees. For example, SOTA systems like NetShare generate UDP packets with TCP flags or DNS packets with out-of-range IPs, severely undermining user trust. While recent systems like NetDiffusion~\cite{jiang2024netdiffusion} attempt post-generation corrections to preserve semantics, they still commit glaring errors. As detailed in \S\ref{rule-enforce}, we found that NetDiffusion frequently breaks sequence number continuity and mishandles TCP handshakes despite its explicit post-hoc constraints.

\myitem{Telemetry imputation.} Recent efforts improve software-based network monitoring by recovering fine-grained time series from coarse-grained telemetry~\cite{gong2024zoom2net}. This is possible because concurrent network signals (\eg queue lengths, ECN marks, and retransmissions) act as correlated symptoms of underlying states like congestion. Zoom2Net~\cite{gong2024zoom2net} leverages generative models to learn these correlations and impute missing details. To improve trustworthiness, it incorporates domain knowledge via manually crafted symbolic rules. However, Zoom2Net's pipeline has two major drawbacks. First, because rules are embedded in the training loss function, the model requires complete retraining for even minor rule adjustments, such as changing the input collection granularity. Second, its reliance on a small, manually curated rule set leaves room for basic errors. For example, when trained on Meta's datacenter dataset to impute ingress bytes per millisecond, Zoom2Net occasionally generates ingress byte counts that are impossibly smaller than the retransmitted bytes in the same interval (\S\ref{rule-enforce}).

\subsection{Pitfalls in Generative ML for Networks} \label{sec:limitations}
Existing approaches fail to integrate ML and formal reasoning for three key reasons:

\myitem{End-to-end reliance.} Driven by advancements like the attention mechanism~\cite{vaswani2017attention}, designers frequently use a single model to learn all correlations directly from data. While minimizing manual effort, this reliance makes systems \emph{uncontrollable} and \emph{brittle}. For instance, operators cannot instruct NetShare to obey basic handshake semantics and must entirely retrain Zoom2Net simply to swap an input signal.

\myitem{Limited rule coverage.} Systems that do incorporate domain-specific rules typically include only a fraction of the necessary constraints. First, they rely on manually specifying rules, which will inevitably lead to poor coverage. Furthermore, many critical networking rules are either non-differentiable or too complex to encode in standard ML loss functions, strictly limiting systems like Zoom2Net to differentiable rules.

\myitem{Knowledge competing with ML.} Rules are frequently applied post-hoc to ``correct'' outputs, placing symbolic knowledge in direct competition with the ML model. This forces a binary choice between trusting the model or the rules. In NetDiffusion~\cite{jiang2024netdiffusion}, for example, if post-processing cannot enforce a rule within a few iterations, the system defaults to the model's output, inevitably yielding data that violates fundamental networking principles.

\begin{figure}[t]
  \centering
  \begin{adjustbox}{max width=\linewidth}
    \begin{subfigure}{.65\linewidth}
      \centering
      \includegraphics[width=\linewidth]{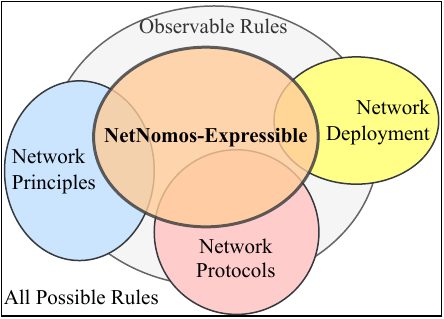}
      \caption{}
      \label{fig:rule_venn}
    \end{subfigure}
    
    \begin{subfigure}{.32\linewidth}
      \centering
      \includegraphics[width=\linewidth]{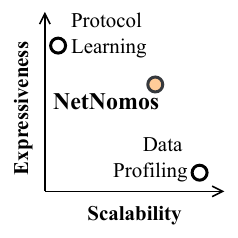}
      \caption{}
      \label{fig:solution_space}
    \end{subfigure}
  \end{adjustbox}
  \caption{(a): \sys finds rules that connect observable variables; these can stem from network principles, protocols, deployment decisions, or their combination. (b): \sys strikes a delicate balance between expressiveness and scalability in the trade-off space for learning network rules.}
\end{figure}

\begin{figure*}[th]
    \centering
    \begin{adjustbox}{width=1\linewidth,center=0pt}
    \includegraphics[width=\linewidth]{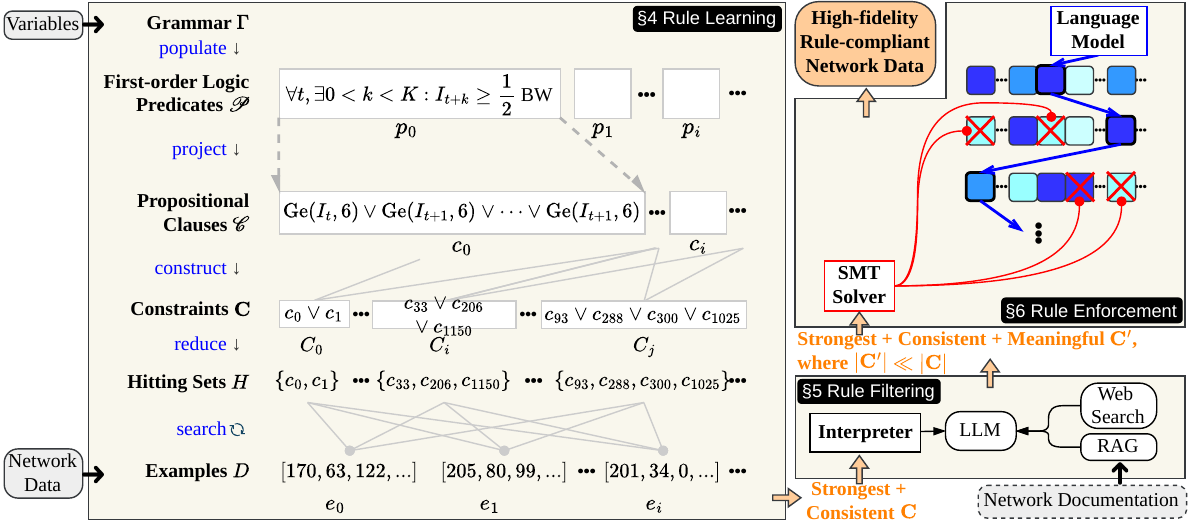}
    \end{adjustbox}
\caption{\sys consists of three stages: \textbf{Rule Learning}, where \sys identifies the minimum set of constraints that are consistent with data (\ie are consistent and strongest) after reducing the problem into the minimum hitting set problem; \textbf{Rule Filtering}, where an LLM (or a human) filters out some of the learned rules as meaningless; and \textbf{Rule Enforcement}: where an SMT solver enforces rules during the token-by-token generation of a language model by invalidating the tokens that if selected by the LM would result in an inconsistent output (\eg a sequence of packets with header fields that violate protocol rules or an imputed fine-grained vector of measurements that defy network principles). }
\label{fig:overview}
\end{figure*}

\section{Overview}

Motivated by the potential of infusing domain knowledge into ML for networking (a neurosymbolic approach), and informed by a clearer understanding of the pain points that deter designers from effectively integrating symbolic and neural components, our vision is to systematize this integration. The goal of this paper is to lay the foundations of a framework that automates rule discovery and seamlessly integrates these rules with generative models. Such a framework will empower network operators to  \emph{(i)} leverage the rich domain knowledge associated with networks in their systems because codifying it will not be a labor-intensive task; \emph{(ii)} benefit from generative models while maintaining their control (for example, they can always add a rule); and \emph{(iii)} avoid choosing between the correctness guarantees offered by domain knowledge (rules) and the predictive capability offered by ML.
Next, we highlight the key insights that allow \sys to realize this vision.



\myitem{\sys mines rules directly from network data.}
We observe that the network itself can be seen as a data-generating process constrained by underlying principles and rules (e.g., capacity limits, routing behavior, protocols) where measurements, packet traces, and other monitoring vectors are feasible solutions of these constraints~\cite{jiang2024netdiffusion,gong2024zoom2net,jin2024pants}. 
Hence, \sys reframes the problem of formalizing network knowledge into logic constraints from a manual, and error-prone task, to a constraint learning problem. At a high level, \sys analyzes network data (\eg measurements, packet headers) and infers rules (formulas over observable fields from data as variables). 
Observable variables (\eg measurements, packet headers) obey rules that stem from protocol definition (\eg TCP handshake), principles (\eg packet drops happen after queue build-up), and deployment decisions (\eg vantage point locations), but also from their combinations.  These are the exact rules that \sys aims to find, as also illustrated in Fig.~\ref{fig:rule_venn}. Observe that \sys cannot find all rules contained in  RFCs or textbooks because they might connect variables that are not in the data, \ie are not measured. Conversely, many of the rules \sys can find and stream from interactions across components cannot be mined from text, which typically explains concepts in isolation. 
By design, \sys produces rules that are easier to use for guiding ML models.
These models are already limited to observable variables during training and are often tailored to specific deployments.

\myitem{\sys reduces rule learning to the minimal hitting set problem.}
Treating the network as a constrained generation process introduces a trade-off between expressiveness and scalability.
Capturing complex relationships requires logical expressiveness, but greater expressiveness enlarges the search space and hurts scalability.
For instance, while most network rules can be expressed in first-order logic (FOL), learning arbitrary FOL rules is undecidable~\cite{boolos2007computability}.

To address this challenge, \sys defines a constraint language $\Gamma$ that restricts expressiveness to a guarded, finite-domain fragment of first-order logic (FOL).
This fragment is expressive enough to capture useful rules, yet structured so that rule learning reduces to a minimal hitting set problem~\cite{reiter1987minhitset} (Fig.~\ref{fig:overview}).
Concretely, $\Gamma$ ranges over a finite set of dataset variables with finite domains (\eg fixed-bitwidth header fields, categorical fields, and bounded counters).
Since all quantification in $\Gamma$ is over this finite universe, each $\Gamma$-constraint can be grounded exactly by enumeration, yielding a propositional formula whose literals correspond to ground predicate instances (including ground arithmetic comparisons over bounded domains).
This grounding is information-preserving under the finite-domain semantics, so satisfaction of a $\Gamma$-formula on a record is unchanged.

The resulting propositional clauses are combined into candidate constraints \textbf{C} as disjuncts.
\sys mutates these candidates (\eg by adding or removing clauses) and retains those consistent with the dataset while concise.
We solve the search as a minimal hitting set problem (\S\ref{subsec:learning}), followed by semantic filtering (\S\ref{sec:semantic-filter}).

\myitem{\sys interjects an SMT solver into the token-by-token generation of a language model (LM).}  
Instead of using knowledge for training or post-inference, \sys includes a true constraint
solver that intersects the LM’s token-by-token inference to guide it towards rule-compliant generation as shown in Fig.~\ref{fig:overview} top right. Tokens are illustrated as blue squares, with color intensity reflecting the probability assigned by the LM. Before each token is selected according to these probabilities by the model, the solver dynamically computes the set of valid next tokens based on the logic rules learned in earlier \sys steps and the sequence of tokens generated so far.
This approach enables easy repurposing of LMs by modifying the rules that are applied during inference rather than retraining or fine-tuning. It also allows network operators to focus on defining useful rules without worrying whether they are differentiable, appear enough in training, or can be embedded in prompts. Finally, \sys enforces rules during inference in a minimally invasive way and preserves the statistical fidelity of ML-learned data distribution. The process of enforcing rules is explained in \S\ref{rule-enforce}.

\subsection{End-to-end View of \sys}
 To better understand \sys end-to-end, let's consider an operator seeking to recover fine-grained (1ms) granularity ingress byte counts of server ports using coarse-grained (50ms) timeseries of congested, retransmitted, dropped, ingress, and egress packet counts. This is the imputation use case described in \S\ref{sec:usecase}. Instead of using Zoom2Net~\cite{gong2024zoom2net}, which requires the operator to manually write rules, the operator uses the first stage of \sys with two inputs: Meta's dataset~\cite{ghabashneh2022millisampler} and the set of variables among the observable ones (those in the dataset) that they believe are correlated. \sys first stage computes a set of non-redundant (later defined as strongest) constraints that are consistent with that dataset $D$. Let us assume that \sys first stage finds the following rules:
 \begin{align}
    &\forall t, 0\leq k<K:\ 0 \le I_{t+k} \le \text{BW}, \tag{\text{R0}} \label{r0} \\
    &\forall t: \texttt{Ingress}_t^K = \sum_{k=1}^{K-1} I_{t+k}, \tag{\text{R1}} \label{r1} \\
    &\forall t: \texttt{Congestion}_t^K > 0 \implies \exists 0\leq k < K: I_{t+k} \geq \frac{1}{2} \text{BW}, \tag{\text{R2}} \label{r2}\\
    &\forall t: \texttt{Connections}_k^K > \text{TH} \implies \nonumber \\
        &\qquad\qquad\qquad(\texttt{Congestion}_t^K > 0 \lor \texttt{InRxmit}_t^K > 0), \tag{\text{R3}} \label{r3} \\
    &\forall t: \texttt{Egress}_t^K > \texttt{InRxmit}_t^K + \texttt{OutRxmit}_t^K, \tag{\text{R4}} \label{r4}
\end{align}
where BW stands for link bandwidth and TH stands for connection count threshold for incast.

 In the second stage, an LLM will filter out meaningless rules in our example,  \ref{r4}.
The remaining constraints $\textbf{C}'$ are passed to an SMT solver, which computes the set of invalid tokens before each new layer of token generation. At each step, the SMT solver incorporates the tokens already generated into the constraint problem at hand. The output of \sys is an input Ingress byte sequence that follows logic rules enforced by the SMT solver and statistical properties enforced by the LM. Visually, \sys aims to constrain the generation to a region that is subject to learned rules (\ie shaded region) in Fig.~\ref{fig:inference_trace}. Note that the ground truth is always within the shaded region, whereas Zoom2Net's output is not.


\section{Extracting Knowledge from Network Data}

\subsection{Formulation of Constraint Modeling} \label{subsec:problem}
We view the network as a data-generating process constrained by rules arising from three sources: network principles, protocols, and deployment specifics. 
Each rule is expressed in formal logic as a symbolic \textit{constraint} $C$. The network data $D$ thus consists of examples that satisfy these constraints.\footnote{A constraint $C$ is a purely syntactic object. We use the term \textit{rule} $R$ to refer to its semantics (either meaningful or meaningless). A meaningful rule must be consistent with the data, but not every consistent rule is meaningful.} An example $e \in D$ depends on the data type: a flow record in NetFlow traces, a sequence of packet headers in PCAP traces, or a time series in performance measurements.
$C$ is \textit{consistent} if it is satisfied by all examples: $D \models C$.

Given a set $\mathbf{C}$ of consistent constraints, each constraint corresponds to a model set $\mathcal{M}(C) = \{ e \in D: e \models C \}$. A constraint $C$ is stronger than $C'$ if $\mathcal{M}(C) \subset \mathcal{M}(C')$, and two constraints $C, C'$ are equivalent if $\mathcal{M}(C) = \mathcal{M}(C')$. A constraint is in its strongest form if there is no non-equivalent $C' \in \mathbf{C}$ such that $\mathcal{M}(C') \subset \mathcal{M}(C)$. Formally, $C$ is strongest iff $\forall C' \in \mathbf{C},\ C' \not\equiv C \implies \mathcal{M}(C) \subset \mathcal{M}(C')$.  

A constraint $C$ is redundant if there exists some $C' \in \mathbf{C}$ with $C' \models C$, in which case $C$ adds no new information beyond what is already entailed by $C'$. Equivalently, redundancy arises whenever $\mathcal{M}(C) \supseteq \mathcal{M}(C')$. Therefore, a constraint is non-redundant precisely when it is maximally strong: it cannot be replaced by a strictly stronger consistent constraint while preserving equivalence.  

Each $C$ captures relationships among fields in $D$, which we treat as \textit{variables} $V$, such as, source/destination IPs and ports, frame/packet/segment sizes, or ECN marked bytes in collected measurements.
Relationships among $V$ can be complex, as they arise from various sources shown in Fig.~\ref{fig:rule_venn}.
Capturing these intricate relationships in a concise and straightforward way requires $C$ to be expressed in first-order logic (FOL).

\mypar{Goal}
Given a network dataset $D$ and its variables of interest $V$, we aim to learn a \textit{minimal constraint theory}, defined as $\text{Th}(\mathbf{C}) \assign \bigwedge_{C \in \mathbf{C}^'} C$, where $\mathbf{C}^' \subseteq \mathbf{C}$ is the subset of consistent and strongest constraints. 
This formulation ensures that $\text{Th}(\mathbf{C})$ avoids thousands of weaker reformulations of the same constraint, which contribute no additional information.

\mypar{Complexity}
Learning FOL formulas from $D$ is an \textit{undecidable} problem, and therefore, one has to restrict FOL to a decidable fragment, \eg the Bernays-Schönfinkel class~\cite{ramsey1987epr}.
Learning a minimal constraint theory even within a decidable fragment is NP-complete~\cite{dershowitz2006scalable,ignatiev2015smallest,nadel2010boosting}.
Consequently, the effectiveness of a rule-leaning method comes down to the efficiency of its search algorithm operating in a huge combinatorial space.

\subsection{Expressive Grammar for Network Data}
\begin{table}[t]
 \centering
 \begin{adjustbox}{width=1\linewidth,center=0pt}
 \begin{tabular}{ lcl } 
  $\langle\texttt{type}\rangle$ & $\mathrel{\mathop:}\assign$ & $\tau \in \{\text{TIME}, \text{SIZE}, \text{ID}, \text{FLAG}, \text{COUNT}\}$ \\[4pt]
  $\langle\texttt{variable}\rangle$ & $\mathrel{\mathop:}\assign$ & $v^{\langle\texttt{type}\rangle} \in V$ \\
  $\langle\texttt{index}\rangle$ & $\mathrel{\mathop:}\assign$ & $0\ |\ 1\ |\ \dots\ |\ K-1$ \\
  $\langle\texttt{svar}\rangle$ & $\mathrel{\mathop:}\assign$ & $\langle\texttt{variable}\rangle^{\{\langle\texttt{index}\rangle\}}$ \\
  $\langle\texttt{context}\rangle$ & $\mathrel{\mathop:}\assign$ & 
     $\{\langle\texttt{svar}\rangle\ (,\ \langle\texttt{svar}\rangle)^{*}\}$ \\[4pt]
  $\langle\texttt{constant}\rangle$ & $\mathrel{\mathop:}\assign$ & $\mathcal{D}(V)\ \cup\ \mathcal{B}(D)$ \\
  $\langle\texttt{operator}\rangle$ & $\mathrel{\mathop:}\assign$ & 
  $+\ |\ \times\ |\ <\ |\ >\ |\ \le\ |\ \ge\ |\ =\ |\ \neq $ \\
  $\langle\texttt{connective}\rangle$ & $\mathrel{\mathop:}\assign$ & $\lor\ |\ \land\ |\ \implies$ \\[4pt]
  \rowcolor{Gainsboro!30}
  $\langle\texttt{aggregator}\rangle$ & $\mathrel{\mathop:}\assign$ & $\max\ |\ \min\ |\ \sum\ |\ \text{avg}$ \\
  \rowcolor{Gainsboro!30}
  $\langle\texttt{avar}\rangle$ & $\mathrel{\mathop:}\assign$ & 
     $\langle\texttt{aggregator}\rangle(\langle\texttt{svar}\rangle^{+})$ \\[4pt]
  $\langle\texttt{lhs}\rangle$ & $\mathrel{\mathop:}\assign$ & 
     $[\langle\texttt{constant}\rangle\langle\texttt{operator}\rangle]\langle\texttt{svar}\rangle^{\langle\texttt{type}\rangle}$ \\
  $\langle\texttt{term}\rangle$ & $\mathrel{\mathop:}\assign$ & 
     $\langle\texttt{constant}\rangle$ \\
     & $|$ & $[\langle\texttt{constant}\rangle\langle\texttt{operator}\rangle]\langle\texttt{svar}\rangle^{\langle\texttt{type}\rangle}$ \\
  $\langle\texttt{rhs}\rangle$ & $\mathrel{\mathop:}\assign$ & 
    $\langle\texttt{term}\rangle\ |\ [\langle\texttt{constant}\rangle\langle\texttt{operator}\rangle]\langle\texttt{avar}\rangle$ \\
  $\langle\texttt{predicate}\rangle$ & $\mathrel{\mathop:}\assign$ & 
     $(\ \langle\texttt{lhs}\rangle^{\tau}\ \langle\texttt{operator}\rangle\ \langle\texttt{rhs}\rangle^{\tau}\ )$ \\[2pt]
  $\langle\texttt{pred\_lit}\rangle$ & $\mathrel{\mathop:}\assign$ & $[\lnot]\langle\texttt{predicate}\rangle$ \\
  $\langle\texttt{predicates}\rangle$ & $\mathrel{\mathop:}\assign$ &
     $\langle\texttt{pred\_lit}\rangle\ (\ \langle\texttt{connective}\rangle\ \langle\texttt{pred\_lit}\rangle\ )^{*}$ \\[6pt]
  $\langle\texttt{qf}\rangle$ & $\mathrel{\mathop:}\assign$ &
   $\langle\texttt{predicates}\rangle$ \\[2pt]
$\langle\texttt{qblock}\rangle$ & $\mathrel{\mathop:}\assign$ &
   $\epsilon$ \\
   & $|$ &
   $\forall\ \langle\texttt{context}\rangle\ |\ \forall\ \langle\texttt{context}\rangle\ \exists\ \langle\texttt{svar}\rangle$ \\[2pt]
$\langle\texttt{constraint}\rangle$ & $\mathrel{\mathop:}\assign$ &
   $\langle\texttt{qblock}\rangle\ :\ \langle\texttt{qf}\rangle$ \\
 \bottomrule
 \end{tabular}
 \end{adjustbox}
 \caption{Grammar $\Gamma$ for the phrase structure of network rules with context window $K$, and user-defined aggregations (shaded).
 Well-formedness of $\Gamma$ requires that variables in $\langle\texttt{lhs}\rangle$ and $\langle\texttt{rhs}\rangle$ of a predicate share the same type $\tau$.}
 \label{tab:grammar}
 \vspace{-4pt}
\end{table}

\mypar{Rich expressiveness}
Within the decidable fragment of FOL, we define the grammar $\Gamma$ such that it is sufficiently expressive to capture most network rules observable from data.

Table~\ref{tab:grammar} summarizes the phrase structures of $\Gamma$.
It supports arbitrary combinations of variables and constants as terms, together with a variety of arithmetic operations.
While being flexible, $\Gamma$ tags every $v \in V$ with one of five types and enforces type checking on predicates.
This typing avoids meaningless constraint candidates, for example, $\texttt{FlowBytes} > \texttt{Duration}$, where \texttt{FlowBytes} is of type COUNT while \texttt{Duration} is of type TIME.
Constants in $\Gamma$ come from two sources:
(1) known variable domains $\mathcal{D}(V)$ derived from existing documentation/specifications, and
(2) background knowledge $\mathcal{B}_{D}$ obtained via profiling.
By default, $\Gamma$ includes a set of well-known constants (\eg ports, protocols, valid combinations of TCP flag bits).
For variables without predefined domains, \sys extracts constants directly from $D$: it selects the top-10 most frequent values for discrete $V$, and for continuous $V$, the five quartiles plus the 90th percentile (p90).
Finally, $\Gamma$ supports user-defined functions such as numerical aggregations.
During evaluation on a concrete record (or context window), we treat these functions as uninterpreted operators and compute their outputs directly, without applying symbolic rewriting or additional constraints.

\mypar{Restrictions}
Without restrictions, a constraint grammar leads to an unbounded search space of candidate formulas~\cite{cohen2006langcomplexity}.
Even with bounded arity $a$ (number of variables), the predicate space $\mathscr{P}$ grows as $|\mathscr{P}| = \mathcal{O}(r|V|^2)$, where $r$ is the number of operators.
The corresponding formula space $\mathscr{F}$ then explodes doubly exponentially, $|\mathscr{F}| = 2^{2^{|\mathscr{P}|}}$.
An overly general grammar design leads to such intractable complexity.
Therefore, it is imperative to specialize $\Gamma$ by enforcing restrictions on its phrase structure, so that the search space becomes tractable while still expressive enough to capture most observable network rules.

$\Gamma$ imposes two main restrictions on constraint candidates.
First, predicates may only contain terms with a single variable on one side of the operator and a single or aggregated variable on the other side.
We observe that such this form of predicates is sufficient in capturing most relationships we can observe from four diverse network datasets (Table~\ref{tab:datasets}, \ref{tab:pcap_bench}--\ref{tab:meta_bench}).
This restriction limits the size of the predicate space ($|\mathscr{P}|$), which in turn bounds the combinatorial search space of all possible constraints, \ie $\mathcal{O}(2^{|\mathscr{P}|})$.
Moreover, it confines the computation domain to Linear Rational Arithmetic (LRA), which keeps $\Gamma$ decidable~\cite{bradley2007calculus}.
Second, the quantifier $\exists$ may only appear after $\forall$ within the context size $K$.
In other words, $\Gamma$ disallows global $\exists$ quantification; instead, $\exists$ is only effective within $K$ consecutive observations.
We impose this restriction for two reasons:
(1) evaluating $\exists$ is prohibitively expensive, even within a limited time window (\eg a protocol execution~\cite{yao2022duoai,hance2021swiss,yao2021distai}); and
(2) network data $D$ typically contains millions of records and is \textit{unbounded in time} (\ie does not have the notion of a single run/execution), making global existential learning impractical.

We argue that the two restrictions do not fundamentally undermine the expressiveness of \sys.
In practice, the wide variety of predicates ($|\mathscr{P}|$) supported by $\Gamma$ still reaches the thousands even with restrictions, whereas prior work~\cite{yao2021distai,ma2019i4,chu2013fastdc,pena2022fastdc++,koenig2020folic3} supports fewer than 100 predicates.
Moreover, while global existential properties are valuable, network operators rarely analyze them in practice since doing so requires substantial compute and/or memory~\cite{hsieh2024netvigil,ghabashneh2022millisampler,li2024reasoning,zhang2022differential,gember2015management,yu2013software}.

\subsection{Limitations of Existing Work}

Existing methods for learning logical rules fall into two categories: data profiling and protocol invariant learning.
The former lacks expressiveness, while the latter fails to scale to the complexity of network data.

Decades of work in data profiling (\eg association rule learning~\cite{pei2007hmine,zhang2002association,wang2002top}, functional dependency discovery~\cite{papenbrock2015functional,papenbrock2016hybrid}) has produced methods with limited formal grammars.
These methods cannot capture the diversity of network rules.
As shown in Fig.~\ref{fig:expressiveness}, their expressiveness bounds the rules they can learn, preventing full coverage of network behaviors.

Protocol learning methods (\eg I4~\cite{ma2019i4}, FOL-IC3~\cite{koenig2020folic3}, DistAI~\cite{yao2021distai}, SWISS~\cite{hance2021swiss}, DuoAI~\cite{yao2022duoai}, Basilisk~\cite{zhang2025basilisk}) are expressive but unsuitable for network data.
They face three fundamental limitations.
{First}, they assume protocol models are \textit{known} and use these as ground truth with verifiers like IVy~\cite{padon2016ivy}.
\sys, in contrast, has no model of the entire network.
{Second}, they rely on \textit{direct} input-output traces from simulation.
\sys operates in a passive setting, learning only from collected traces.
Without interactive feedback, learning an equivalent automaton is provably infeasible~\cite{angluin1987lstar}.
{Third}, they are designed for learning one protocol at a time.
This assumption restricts their predicate space $\mathscr{P}$ to variables of a single entity, keeping $|\mathscr{P}|$ small.
Network data $D$ is heterogeneous, containing many entities and their interactions.
As a result, \sys’s predicate space is $2$--$15\times$ larger, and since the search space is exponential in $|\mathscr{P}|$, scalability becomes the key barrier.
As shown in Fig.~\ref{fig:scalability}, a SOTA protocol learning method learns only a fraction of benchmark rules, while \sys learns nearly all within the same time budget.
\subsection{From  Constraints to Hitting Sets}

Prior work enumerates the formula space $\mathscr{F}$.
This approach does not scale for two fundamental reasons.
\textit{First}, evaluating the strength of each candidate constraint requires scanning \textit{all} examples.
The cost is $\mathcal{O}(|D|\cdot 2^{|\mathscr{P}|})$.
In networking, $|D|$ is often in the millions.
\textit{Second}, instantiating all constraints at the same strength causes \textit{exponential level-wise growth}.
Weakening produces more candidates at the next level than the current level provides.
For example, if $P_1 \lor P_2$ is too strong (where $P_i$ are predicates), weakening by adding $P_3$ yields $|\mathscr{P}|$ new candidates to evaluate.

\sys takes an indirect route, reducing the search for strongest constraints to a Minimal Hitting Set problem~\cite{reiter1987minhitset,chu2013fastdc}.
We design $\Gamma$ so that all variables range over finite domains (\eg bounded counters, finite-width header fields, and a finite context window $K$).
Under these finite-domain semantics, any $\Gamma$-constraint can be grounded exactly by enumerating quantified variables, yielding a finite Boolean formula whose literals are ground predicate instances (including ground arithmetic comparisons).
This grounding is information-preserving for $\Gamma$ and enables solving the search via hitting set.
Thus every constraint $C$ can be represented as a set of \textit{clauses} in propositional logic, \eg $C = (c_0 \lor c_1 \lor c_2) \equiv \{c_0, c_1, c_2\}$.
A \textit{clause} is composed of one or more propositions, \eg protocol is TCP: $\text{Eq}(\texttt{Proto}, \textsl{TCP})$.
Fewer clauses mean a stronger constraint: $C' = \{c_0, c_2\}$ is stronger than $C$, \ie $C' \vdash C$.
Each clause $c$ satisfies a subset of examples.
Define its evidence set as $E_c \assign \bigcup_e^D e \models c$.
All clauses in a consistent $C$ together must satisfy $D$.
Equivalently, $C \models D \Longleftrightarrow \left(\bigcup_{c}^{C} E_c = D\right)$.

Intuitively, learning a consistent constraint means finding clauses whose evidence sets include $D$.
This process reduces to the Hitting Set problem: given a set of clauses, find a subset $H$ of size at most $s$ whose evidence sets together “hit” all examples in $D$.
The bound $s$ limits the number of clauses a constraint may contain.
\begin{txtframe}
\begin{thm}\label{thm:hitset}
Learning a consistent constraint $C$ on examples $D$ is equivalent to finding a hitting set $H$ of clauses whose evidence sets hit $D$.\footnotemark 
\end{thm}
\end{txtframe}\footnotetext{Proof sketch: Appendix~\ref{thm:hitset}.}

A hitting set $H$ is \textit{minimal} if no proper subset of $H$ is also a hitting set.
In that case, $C \assign \bigwedge_{c}^{H} c$ uses the fewest possible clauses and is therefore the strongest:
\begin{txtframe}
\begin{lem}\label{thm:minhitset}
A minimal hitting set corresponds to a constraint that is the strongest.
\end{lem}
\end{txtframe}

This reduction gives \sys two scaling advantages.
{First}, it avoids exhaustive per-candidate evaluation on $D$.
As explained in \S\ref{subsec:learning}, \sys scans $D$ once to construct evidence sets, yielding complexity $\mathcal{O}(|D|\cdot|\mathscr{P}|)$ instead of $\mathcal{O}(|D|\cdot 2^{|\mathscr{P}|})$.
{Second}, it traverses the search space without level-by-level instantiation by strength, thereby avoiding exponential search space explosion.
Next, we explain the learning process step by step.







\subsection{Rule Learning Method of \sys} \label{subsec:learning}

\begin{figure*}[t]
  \centering
    
  \begin{minipage}[t]{0.67\textwidth}
    \centering
    \begin{adjustbox}{width=\linewidth,center=0pt}
      \includegraphics{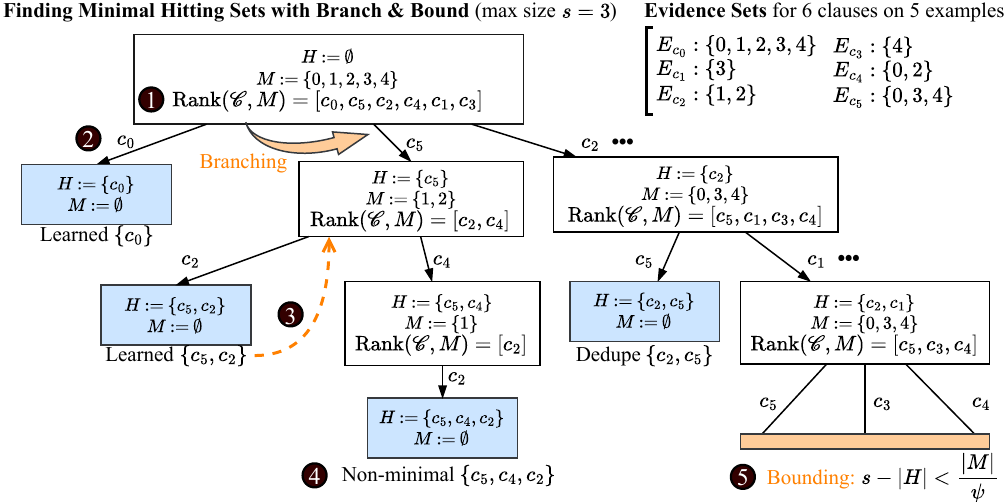}
    \end{adjustbox}
    \caption{\sys learns complex FOL constraints by systematically finding minimal hitting sets.}
    \label{fig:rule_learning}
  \end{minipage}
  \hfill
  \begin{minipage}[t]{0.31\textwidth}
    \centering
    \begin{adjustbox}{width=\linewidth,center=0pt}
      \includegraphics{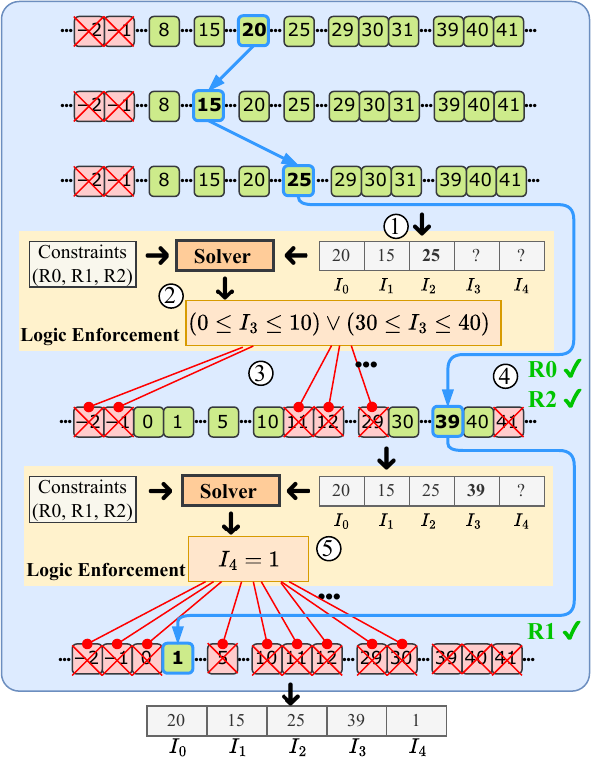}
    \end{adjustbox}
    \caption{\sys invokes a solver during inference to filter out invalid tokens that will cause rule violations.}
    \label{fig:inference}
  \end{minipage}
 \end{figure*}

\mypar{Search space projection}
The first step of the learning process is constructing the search space. 
\sys starts by populating the space $\mathscr{P}$ of FOL predicates that are allowed by $\Gamma$:
$\mathscr{P} \mathrel{\mathop:}= \{ \texttt{Ingress}_t^K = \sum_{k=0}^{K-1} I_{t+k},\ 
    \texttt{Ingress}_t^K > \sum_{k=0}^{K-1} I_{t+k},\  
    \texttt{Congestion}_t^K = 0,\ 
    \texttt{Congestion}_t^K > 0,\ 
    \exists 0 \leq k < K-1: I_{t+k} \geq \frac{1}{2}\text{BW},\dots\nonumber\}.$

Then, \sys propositionalizes $\mathscr{P}$ by projecting all predicates therein to propositional clauses ($\mathscr{C}$).
Specifically, this process unrolls quantifier $\exists$ in FOL and translates aggregation functions to basic propositions:
\begin{align}
    \mathscr{C} &\mathrel{\mathop:}= \{ 
    \text{Eq}(\texttt{Ingress}_t^K, I_t+I_{t+1}+\dots+I_{t+K-1}), \tag{$c_0$}\label{c0}\\
    &\text{Gt}(\texttt{Ingress}_t^K, I_t+I_{t+1}+\dots+I_{t+K-1}),\ \dots\nonumber \\
    &\text{Eq}(\texttt{Congestion}_t^K , 0), \tag{$c_2$}\label{c2}\\
    &\text{Gt}(\texttt{Congestion}_t^K , 0),\ \dots,\nonumber\\
    &\left(\text{Ge}(I_t, 6) \lor \text{Ge}(I_{t+1},6) \lor\dots \lor\text{Ge}(I_{t+K-1},6)\right), \tag{$c_5$}\label{c5}\\\ \dots&\}, \nonumber
\end{align}
where constants are materialized in Gbps for simplicity.
 
This process produces a clause space $\mathscr{C}$, from which constraint formulas are constructed.

\mypar{Building evidence sets}
For each $c \in \mathscr{C}$, \sys builds an evidence set $E_c$ containing indices of all examples satisfying $c$: $\forall i \in E_c: D_i \models c$.
In Fig.~\ref{fig:rule_learning}, six clauses \ref{c0}--\ref{c5} and five examples (0--4) yield six sets $E_0$--$E_5$.
For example, all examples satisfy \ref{c0}, while only $D_3$ satisfies $c_1$.
This step runs in $\mathcal{O}(|D|\cdot|\mathscr{P}|)$.

\mypar{Finding minimal hitting sets}
\sys then solves the minimal hitting set problem using branch and bound (Fig.~\ref{fig:rule_learning}).
Clauses are ranked by \textit{cover}, the number of uncovered examples they hit: $\text{Cover}(c) = \mid E_c \setminus \bigcup_{c'}^{{H}}{E_{c'}}\mid$.
The highest-cover clause becomes the pivot. For example, \ref{c0} is chosen first, producing $H={\text{\ref{c0}}}$ which hits all examples and forms a valid hitting set.
Unhit examples are tracked as $M = D \setminus \bigcup_{c\in H} E_c$.
When $M=\emptyset$, $H$ is valid; otherwise, \sys branches on new pivots.
For instance, branching on \ref{c5} yields $M=\{1,2\}$, and further branching on \ref{c2} produces another hitting set $\{\text{\ref{c5}},\text{\ref{c2}}\}$.
Non-minimal sets, such as $\{\text{\ref{c5}},c_4,\text{\ref{c2}}\}$, are discarded.

\mypar{Bounding search}
To prune, \sys computes the maximum possible cover $\psi=\max_c \text{Cover}(c)$ and stops search when $s-|H| < |M|/\psi$, where $s$ is the maximum formula size.
This optimistic bound prevents exploring subtrees that cannot yield valid sets.

\mypar{Minimal constraint theory \& proof system}
The process yields minimal hitting sets such as $\{\text{\ref{c0}}\}$ and $\{\text{\ref{c5}},\text{\ref{c2}}\}$, which correspond to constraints \ref{r1} and \ref{r2}:
\begin{align*}
    \{\text{\ref{c0}}\} &\mapsto \text{\ref{c0}} \equiv C1 \equiv \ref{r1},\ \text{and}\\
    \{\text{\ref{c5}}, \text{\ref{c2}}\} &\mapsto (\text{\ref{c5}} \lor \text{\ref{c2}}) \equiv (\lnot \text{\ref{c5}} \implies \text{\ref{c2}}) \equiv C2 \equiv \ref{r2}.\\
\end{align*}

\sys then constructs a constraint theory $\text{Th}(\mathbf{C})=\bigwedge_{C \in \mathbf{C}} C$ and supports reasoning with the Fitch proof system~\cite{fitch1963logical}, which is sound and complete for propositional clauses.
Given a query $q$,
\begin{align}
f_{\text{Fitch}}: q \mapsto \left\{\top, \bot,\texttt{?}\right\}=\text{Th}(\mathbf{C})\vdash q,\label{eq:inference}
\end{align}
where $f_{\text{Fitch}}$ returns $\top$ if $q$ is derivable from $\text{Th}(\mathbf{C})$, and $\bot$ if it creates a contradiction, and \texttt{?} if $q$ is contingent.
Thus, all derived constraints are correct, and all entailed constraints are derivable.

\section{Semantic Filtering}
\label{sec:semantic-filter}

A learned constraint can be syntactically valid but semantically meaningless (for example, \ref{r4}).
This occurs because \sys, like all rule-learning methods, reasons about syntax but not meaning.
The problem is severe: thousands or even millions of valid formulas may align with the data~\cite{pena2022fastdc++,ben2024cafa,yao2022duoai,yao2021distai,ma2019i4,hance2021swiss}, and the issue grows with grammar expressiveness.

The consequences are twofold:
(1) reduced interpretability of the learned rules, and
(2) heavy overhead when applying or enforcing them.
Although experts can often spot meaningless rules, manually filtering such large sets is infeasible.

\sys addresses this challenge using large language models (LLMs).\footnote{We use LLM to refer to models with billions of parameters trained on massive corpora, while LM denotes any model over discrete vocabulary.}
While LLMs can hallucinate, the risk here is bounded:
(1) they only filter rules already valid with respect to the data, and
(2) they are tasked with semantic reasoning, not generation, which plays to their strength.

The semantic filter (Fig.~\ref{fig:overview}) takes as input the constraints learned by \sys.
It first interprets raw SMT-LIB formulas into logical expressions with semantic values.
For example,
\texttt{(assert (forall ((e Flow)) (=> (not (= (Proto e) 6)) (= (Flags e) 0))))}
becomes
\texttt{"$\forall e: e.\texttt{Proto} \neq \texttt{TCP} \Rightarrow e.\texttt{Flags} = \emptyset$"}
(Rule \#6, Table~\ref{tab:cidds_bench}).
This can also be translated into plain English, though experiments (Fig.~\ref{fig:llm_filter}) show no added benefit for filtering.
Next, \sys queries an external LLM with the interpreted rules and an instruction to identify semantically meaningful ones.\footnote{This step involves prompt design, which we do not optimize.}
The LLM can also be augmented with lightweight tools such as web search or retrieval-augmented generation (RAG) over RFCs, specifications, and related papers.
Rules marked as meaningful are kept; the rest are discarded.

As shown in Fig.~\ref{fig:llm_filter}, this approach is highly effective.
Leveraging the interpretability of \sys’s rules, the semantic filter removes meaningless rules with over 98\% precision and under 0.73\% false positive rate.
\section{Rule Enforcement}
\label{rule-enforce}

\begin{figure}[t]
    \centering
    \begin{adjustbox}{width=0.9\linewidth,center=0pt}
    \includegraphics[width=\linewidth]{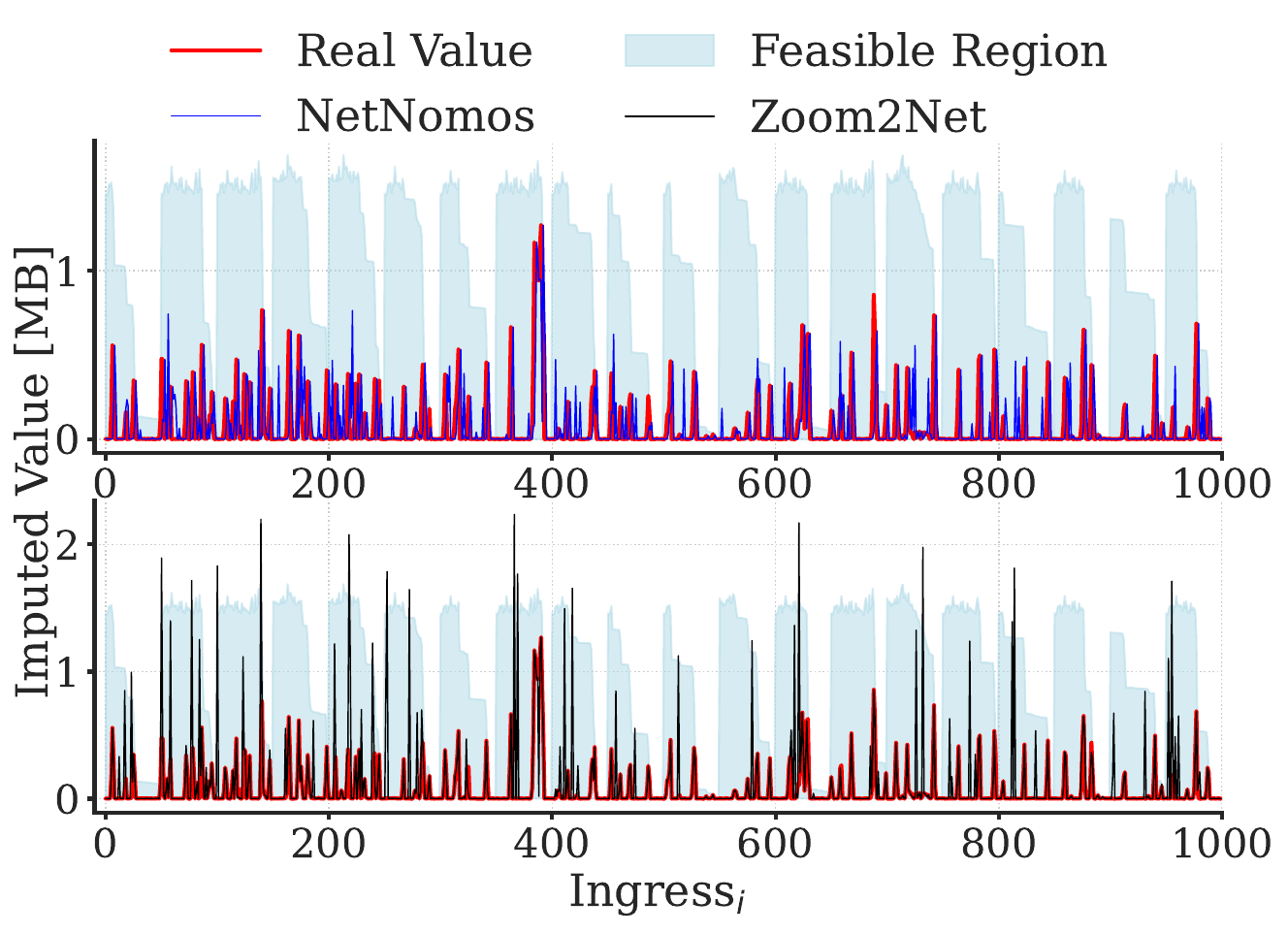}
    \end{adjustbox}
\caption{(Upper) \sys enforces learned rules at inference time, guaranteeing that model outputs remain within the feasible region defined by constraints.
    (Lower) In contrast, outputs of Zoom2Net~\cite{gong2024zoom2net} frequently exceed the boundaries.} 
\label{fig:inference_trace}
\end{figure}

We present \sys’s inference-time rule enforcement.
The method interleaves with the LM’s token-by-token decoding and steers generation toward rule-compliant outputs (Fig.~\ref{fig:inference}).
An SMT solver adds some latency, yet it lets \sys combine neural and symbolic reasoning.
Symbolic reasoning enforces diverse constraints, including arithmetic and non-differentiable ones, without manual rule checks by operators.
At the same time, \sys preserves the benefit of statistical learning by intervening only when needed.

\mypar{Example}
Consider imputing $[I_0,\dots,I_4]$ under constraints \ref{r0}--\ref{r2}.
After the LM emits a complete value (\eg $I_2$ at \circlewhite{1}), \sys calls the solver with all three constraints, instantiated with the values generated so far.
This partial instantiation identifies which constraints are active and what they imply for the next symbol.
If a rule such as \ref{r2} is already satisfied by earlier values, \sys deactivates it when computing the feasible region for $I_3$.
If not, the solver uses all relevant rules to derive the valid range for $I_3$ (\circlewhite{2}).
\sys then prunes any candidate value of $I_3$ that lies outside this region (\circlewhite{3}).
The chosen value (\eg $I_3=39$) is therefore guaranteed to satisfy all constraints (\circlewhite{4}).
With aggregation rules such as \ref{r1}, the feasible region may collapse to a single value, which the LM then emits (\circlewhite{5}).


\myitem{Fine-grained, minimally invasive control.}
A key challenge is granularity mismatch:
The LM generates opaque tokens from a tokenizer, while
the solver reasons over interpretable variables such as ingress bytes or ECN markings.
For example, the solver may require $I_4=6$ to satisfy \ref{r2}, while the LM’s vocabulary may not contain a standalone token \texttt{"6"}.
Instead, the digit may appear only inside subword tokens such as \texttt{"062"} or \texttt{"\textvisiblespace6"}.
This mismatch can force invasive control that harms generation quality~\cite{BeurerKellner2024domino,geng2023gcd,poesia2022synchromesh}.

\textit{\sys resolves this mismatch with per-field vocabularies and character-level control.}
For each variable in the dataset, \sys instantiates a field-specific vocabulary.
For numerical fields (\eg ingress volume, packet count), \sys treats numbers as plain text~\cite{raffel2020t5} and uses character-level tokenization~\cite{tay2021charformer}, generating digits one by one.
For categorical fields (\eg protocols, ports), \sys tokenizes entire values as single units.
For example, \texttt{TCP} is never split into \texttt{T} and \texttt{CP}.
This design gives \sys control that is at least as fine as the solver’s variable granularity.

During inference, \sys builds token transition systems on the fly.
Given solver-derived feasible ranges, it constructs a labeled transition system~\cite{tretmans2008model,classen2012featured,van1994logic} across variables and an unlabeled one within the digits of a numerical variable.
The current state corresponds to the last emitted token, and the next states include all tokens that keep the partial value within the feasible region.
Fig.~\ref{fig:inference_trace} shows this process in action.
\sys keeps imputed values within the valid (shaded) region at every step, while outputs of Zoom2Net~\cite{gong2024zoom2net} frequently exceed the boundaries.

\section{Evaluation} \label{sec:eval}


Our evaluation answers the following questions:

\begin{enumerate}[leftmargin=*, label=E1.\arabic*, ref=E1.\arabic*, itemsep=0pt, topsep=2pt]
    \item\label{e1.1} Can \sys learn complex network rules that require sufficient grammatical expressiveness?
    \item\label{e1.2} Is \sys more scalable than equally expressive SOTA rule-learning methods?
\end{enumerate}
\begin{enumerate}[leftmargin=*, label=E2., ref=E2., itemsep=0pt, topsep=2pt]
    \item\label{e2.1} Can \sys reliably filter out syntactically valid but semantically meaningless rules?
\end{enumerate}
\begin{enumerate}[leftmargin=*, label=E3., ref=E3., itemsep=0pt, topsep=2pt]
    \item\label{e3.2} Can the same GPT-2 model, without retraining or fine-tuning, achieve on-par performance for distinct tasks (imputation, generation, prediction) with SOTA systems tailored to each task, only by enforcing rules at inference?
\end{enumerate}
Testbed setup for the experiments is reported in Appendix~\ref{apdx:testbed}.



\begin{figure}[t]
  \centering
  \begin{subfigure}[b]{0.60\linewidth}
    \centering
    \begin{adjustbox}{width=\linewidth,center}
      \includegraphics{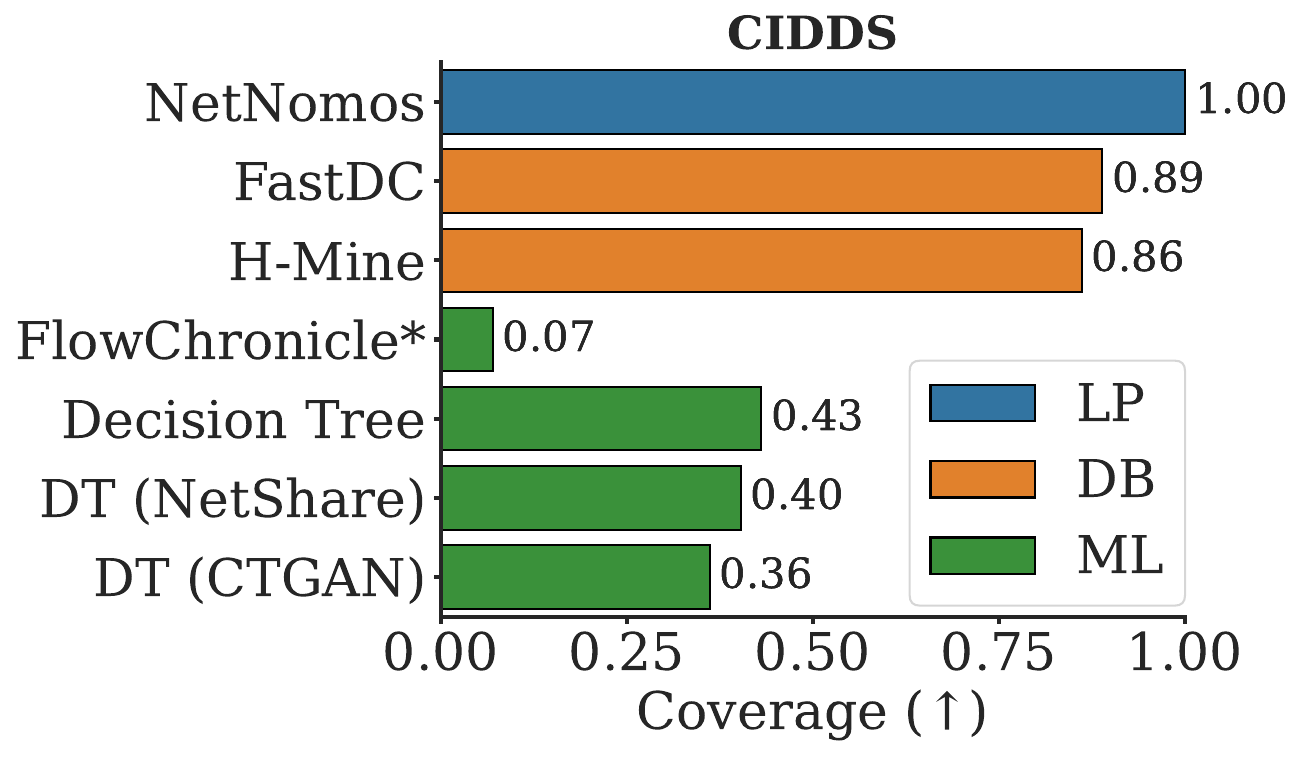}
    \end{adjustbox}
    \caption{}
    \label{fig:expressiveness}
  \end{subfigure}%
  \begin{subfigure}[b]{0.4\linewidth}
    \centering
    \begin{adjustbox}{width=\linewidth,center}
      \includegraphics{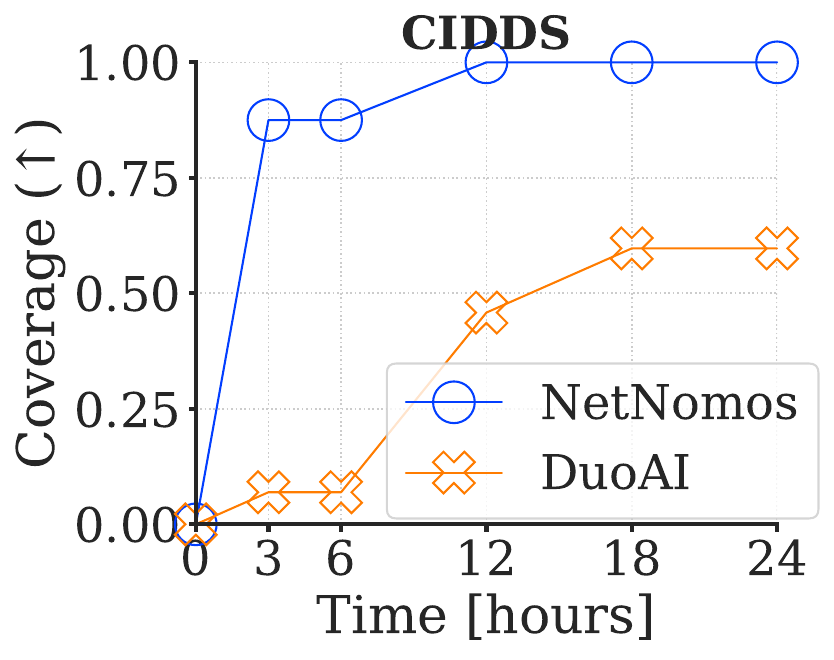}
    \end{adjustbox}
    \caption{}
    \label{fig:scalability}
  \end{subfigure}
  \caption{(a): \sys is much more expressive in capturing network rules than existing SOTA methods from other domains.
Theoretically, $\Gamma$ can express all benchmark rules~(\S\ref{subsec:rulesets}), and its coverage result is bounded by scalability.
(b): \sys outscales DuoAI, achieving $>$75\% rule coverage in half the time and up to 6.5\x better scalability compared to DuoAI. \sys enables practical learning of complex real-world network rules from data within a reasonable time budget. (Full results: Fig.~\ref{fig:expressiveness2}, \ref{fig:scalability2}. Appendix~\ref{apdx:fig-express}, \ref{apdx:fig-scale}).}
  \label{fig:learning}
\end{figure}



\subsection{Benchmark Rulesets} \label{subsec:rulesets}
\mypar{Rule learning rulesets}
To address \ref{e1.1} and \ref{e1.2}, we curate a comprehensive benchmark ruleset for each dataset to compare \sys against baseline rule-learning methods.
We leverage the benchmark rules from prior papers on the corresponding datasets, as well as from RFCs and IANA standards.
As shown in Table~\ref{tab:datasets} in Appendix~\ref{apdx:ds_and_rules}, these rulesets together cover all three types of network rules~(Fig.~\ref{fig:rule_venn}).
We manually translate the rules into formal logic; example rules and their semantics are presented in Tables~\ref{tab:pcap_bench}--\ref{tab:meta_bench} in Appendix~\ref{apdx:expl-rules}.
For all datasets, we set the formula size limit to 12 for \sys.

\mypar{Rule filtering rulesets}
To answer \ref{e2.1}, we build rulesets that contain both meaningful and meaningless rules for evaluating the semantic rule filter of \sys.
Meaningful rules are consistent by definition, while not all consistent rules are meaningful~(\S\ref{subsec:problem}).
We generate meaningless rules from the benchmark rulesets (Table~\ref{tab:datasets}).
The benchmark rules are both consistent and meaningful.
To create meaningless ones, we apply Semantic Fusion~\cite{winterer2020semanticfusion}.
Specifically, we mix the premises and conclusions of logical implications.
For example, given two benchmark rules $X \implies Y$ and $Z \implies Q$, we construct $X \implies Q$ and $Z \implies Y$.
We then evaluate these fused rules on the corresponding datasets.
The ones that remain consistent with the network data are treated as meaningless rules.
This process produces cases such as $\texttt{SrcPt} = 80 \implies \texttt{SrcPt} \neq 433$, which is consistent but meaningless.
Table~\ref{tab:filter_rulesets} in Appendix~\ref{apdx:filter_rulesets} summarizes the resulting rulesets used for evaluating the semantic filter.


\subsection{Rule Learning}

\mypar{Baselines}
We compare \sys with representative rule-learning methods from three domains: logic programming (LP), databases (DB), and machine learning (ML).
Specifically,, we evaluate LP-based  DuoAI~\cite{yao2022duoai}; DB-based FastDC~\cite{chu2013fastdc,pena2022fastdc++} and H-Mine~\cite{pei2007hmine}; ML-based FlowChronicle~\cite{cuppers2024flowchronicle} and Decision tree (DT). Details are provided in Appendix~\ref{apdx:baselines}.

\mypar{Metric}
We evaluate the coverage of all rule-learning methods on the benchmark rulesets~(\S\ref{subsec:rulesets}). We run \sys, as well as baselines, for 24 hours on each dataset.
We then collect their learned rules \textbf{C} and feed them into \sys's proof system (Eqn.~\ref{eq:inference}) for evaluation.
We then query the proof system with each benchmark rule.
A rule is counted as \textit{learned} if the system returns $\top$ (\ie the benchmark rule is derivable from $\text{Th}(\textbf{C})$).
Otherwise, we consider that rule as not learned, \ie the result is $\texttt{?}$ or $\bot$.
Rule coverage is then computed as: $\text{Coverage} = \frac{\text{\# of learned rules}}{\text{Total \# of rules in benchmark}}$.


\smallskip
\mypar{Expressiveness results}
We first evaluate the expressiveness of \sys.
All baselines, except for DuoAI, are limited in expressiveness; that is, their underlying rule-learning methods cannot capture all types of network rules in the benchmarks.
Fig.~\ref{fig:expressiveness} compares \sys against these expressiveness-bounded methods (full results shown in Fig.~\ref{fig:expressiveness2}, Appendix~\ref{apdx:fig-express}).
\sys is able to capture nearly all benchmark rules, missing only 2 rules in the Netflix benchmark and 6 rules in MAWI.
The grammar $\Gamma$ of \sys can, in principle, express all benchmark rules.
The few rules not learned are long and complex protocol rules that exceed the current scalability of \sys.
Improving scalability to cover such rules is an avenue for future work.

\smallskip

\mypar{Scalability results}
Fig.~\ref{fig:scalability} depicts the rule coverage and running time for \sys and DuoAI (full results in Fig.~\ref{fig:scalability2}, Appendix~\ref{apdx:fig-scale}).
Both methods are bound by scalability rather than expressiveness.
Within the first 12 hours, \sys achieves over 75\% coverage on all four benchmarks, while DuoAI reaches at most 60\% coverage after 24 hours.
The advantage of \sys is most pronounced on the Netflix and MAWI benchmarks, where constraint sizes are mostly larger than 8: here, \sys scales 5.7--6.5\x better than DuoAI.
In contrast, the CIDDS and MetaDC benchmarks consist mostly of smaller constraints of size less than 5, where the gap is relatively narrower.



\begin{figure}[t]
    \centering
    \begin{adjustbox}{width=1\linewidth,center=0pt}
    \includegraphics[width=\linewidth]{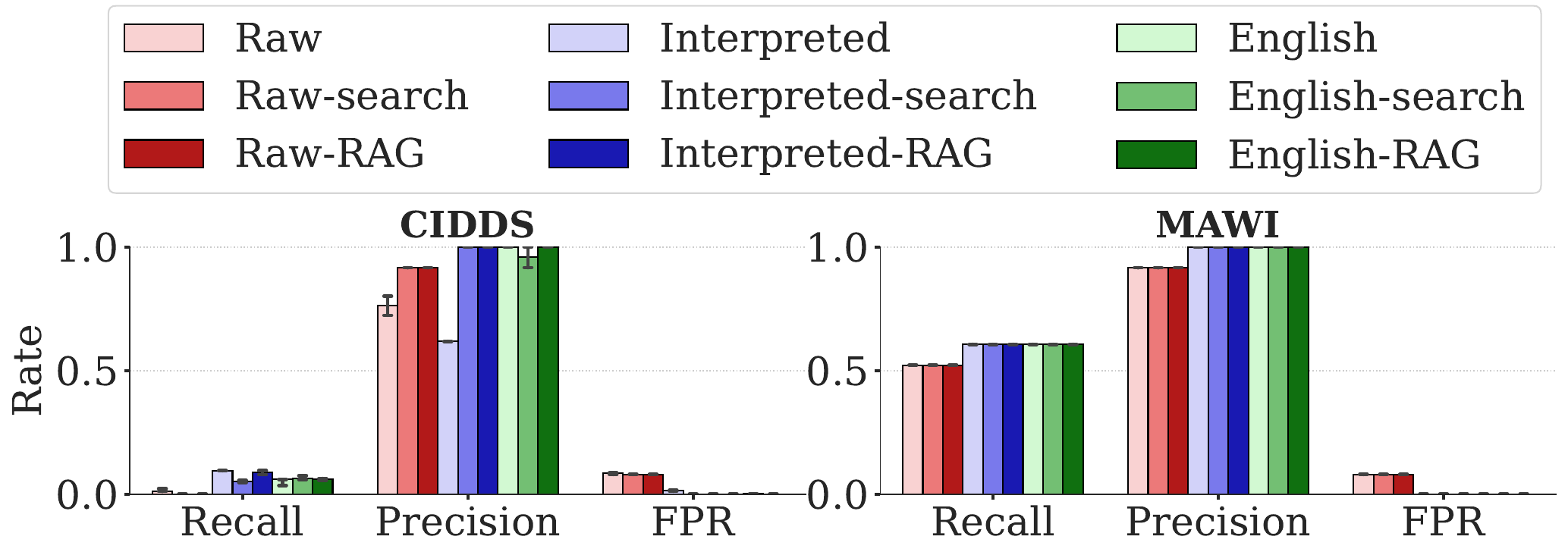}
    \end{adjustbox}
\caption{\sys's filtering stage can effectively avoid carrying meaningless (albeit consistent) rules to the next stage, achieving a 9.2\% FPR. 
\sys also integrates simple interpretation (\texttt{Interpreted}*) and lightweight tools with the LLM such as web search or RAG, which leads to $>$98.1\% precision and $<$0.73\% FPR. (Full results: Fig.~\ref{fig:llm_filter2}, Appendix~\ref{apdx:llm-filter})}
\label{fig:llm_filter}
\end{figure}


\begin{figure*}[ht]
    \centering
    \begin{adjustbox}{width=0.9\linewidth,center=0pt}
    \includegraphics[width=\linewidth]{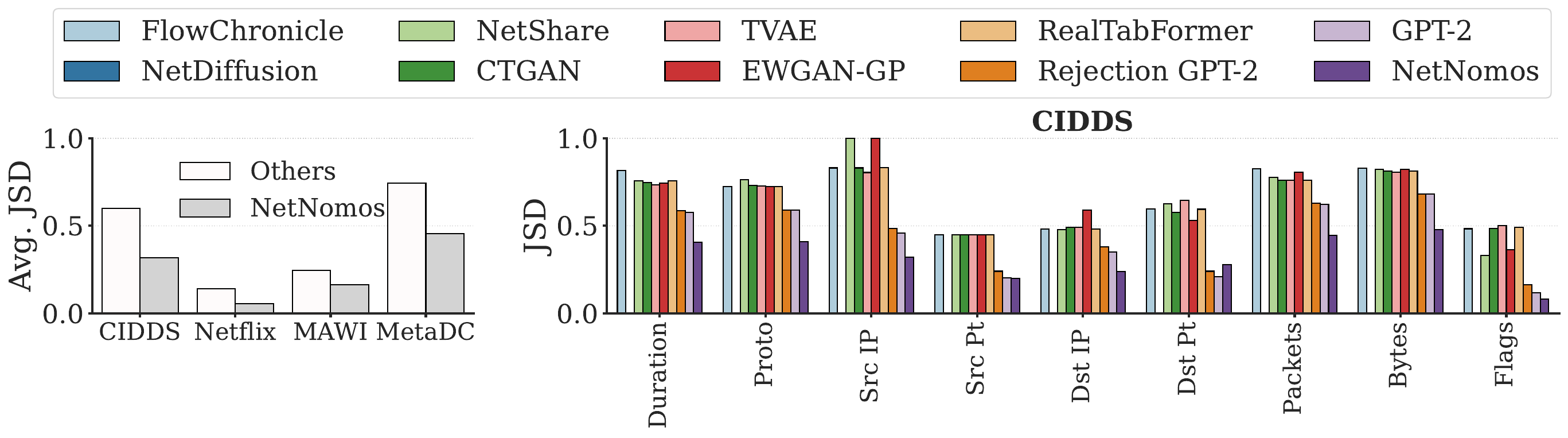}
    \end{adjustbox}
\caption{\small\sys generates synthetic data of higher fidelity, achieving both lower JSD and EMD across four datasets compared to NetShare~\cite{yin2022netshare}, NetDiffusion~\cite{jiang2024netdiffusion}, FlowChronicle~\cite{cuppers2024flowchronicle}, and other generative ML models. For the Netflix and MAWI datasets, the EMD values are truncated to improve the visibility of the lower range. Average (\texttt{Avg.}) JSD and EMD values are computed \textit{without considering the maximum}. (Full results are reported in Fig.~\ref{fig:syngen2}, Appendix~\ref{apdx:syngen}.)}
\label{fig:syngen}
\end{figure*}

\subsection{Rule Filtering}

We use the GPT-4.1 API from OpenAI as the LLM component of \sys's semantic filter.
The API is invoked with a fixed prompt (shown in Appendix~\ref{apdx:prompt}).
Each query appends the rules in one of three forms: \texttt{Raw} formulas in SMT-LIB format, \texttt{Interpreted} formula expressions, or plain \texttt{English} translations.

\mypar{Metrics}
We collect the indices of rules that the LLM marks as meaningful.
From these results, we compute recall (TP / (TP + FN)), precision (TP / (TP + FP)), and false positive rate (FP / (FP + TN)) with the following definitions:
\begin{itemize}[leftmargin=2.5em, nosep, topsep=2pt]
    \item[(TP)] True positives: correctly identified meaningful rules.
    \item[(FP)] False positives: meaningless rules wrongly marked.
    \item[(TN)] True negatives: correctly identified meaningless rules.
\end{itemize}

\mypar{Results}
Fig.~\ref{fig:llm_filter} shows the performance of \sys's semantic filtering (full results in Fig.~\ref{fig:llm_filter2},  Appendix~\ref{apdx:llm-filter}).
Even raw logic formulas (\texttt{Raw}) achieve 82.5\% precision with only 9.2\% FPR.
Interpreting the raw SMT-LIB format as logical formulas with semantic values further improves performance: precision rises by 5.9\% (from 86.4\% to 91.5\%), and FPR drops by 633.6\% (from 6.9\% to 0.82\%).
We then improve filtering by enabling web search and giving the LLM access to a vector database that indexes relevant documents such as protocol RFCs~\cite{rfc1122,rfc4443,rfc6335,rfc7323,rfc7605,rfc768,rfc791,rfc793,rfc8200,rfc9293,rfc9673} and dataset-related papers (\eg \cite{ring2019rulecidds,ring2017cidds,mawi2006traffic,schoen2024rulecidds,jacobs2022trustee}).
With this lightweight tool use, the semantic filter reaches at least 98.1\% precision and at most 0.73\% FPR.
Translating formulas into plain English provides no additional gain, which indicates that \sys’s interpretable rules are already sufficient for reliable selection.
The filtering performance on MetaDC is slightly lower since there is little standard documentation on network principles compared to the abundant protocol specifications.

The main drawback of this approach is recall:
even with interpretation and tool use, recall across all four datasets remains below 76.1\%.
In practice, the LLM is more likely to reject a rule as meaningless, even when given supporting resources.
We argue that this conservative behavior is beneficial.
\sys often learns thousands of syntactically valid rules (Table~\ref{tab:selection_rates} in Appendix~\ref{apdx:filter-rate}).
Enforcing all of them during model inference would put excessive strain on the solver and increase latency.
High precision in filtering therefore reduces overhead and ensures practical deployment.
We acknowledge the recall limitation but improvements in LLM performance are likely to strengthen filtering in the future.



\begin{figure}[t] 
    \centering
    \begin{adjustbox}{width=0.9\linewidth,center=0pt}
    \includegraphics[width=\linewidth]{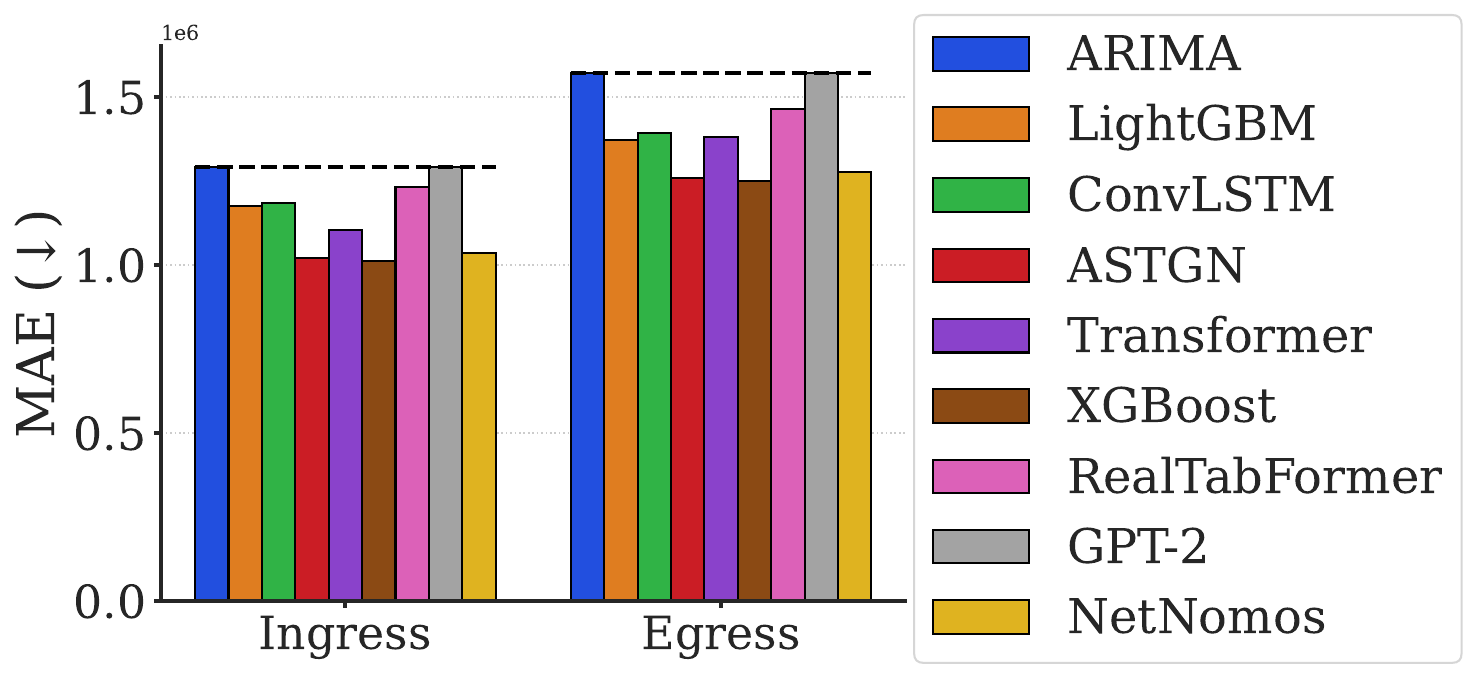}
    \end{adjustbox}
\caption{By enforcing learned rules, \sys is able to improve the generic GPT-2 (marked by dashed line) by 15.04\%--42.67\% across various metrics and achieve competitive performance against regression and the specialized model, ASTGN~\cite{peng2024astgn}. (All six evaluation metrics are reported in Fig.~\ref{fig:forecast2}, Appendix~\ref{apdx:forecast}).}
\label{fig:forecast}
\end{figure}


\section{End-to-end \sys Case Studies} 

For \ref{e3.2}, we evaluate \sys end-to-end on \emph{(i)} data synthesis~(\S\ref{sssec:syngen}); \emph{(ii)} traffic forecasting~(\S\ref{sssec:forecast}); and \emph{(iii)} telemetry imputation~(\S\ref{sssec:impute}). 

\mypar{\sys uses GPT-2}
\sys is LM-agnostic and can naturally benefit from a more powerful model.
Still, to ensure that improvements come from rule learning and enforcement rather than the inherent strength of the LM, we deliberately choose a generic, weaker LM: GPT-2~\cite{radford2019gpt2}.

\mypar{\sys tailors the GPT-2 per task by inference rules}
We use the \textit{same} GPT-2 model, trained once on the network data synthesis task, for all three use cases without retraining or fine-tuning.
What varies across tasks is the set of rules enforced by \sys to contain those that involve task-related variables.
For data synthesis, all rules are applied because every field is generated.
We acknowledge an approximately 5\x inference-time efficiency cost from enforcing \sys rules~\cite{he2025lejit}, due to the overhead of validating the LM's decisions with an SMT solver.
We leave performance optimization to future work.


\subsection{Network Data Synthesis} \label{sssec:syngen}

\mypar{Baselines}
We evaluate \sys against three network-specific generators: NetShare~\cite{yin2022netshare}, NetDiffusion~\cite{jiang2024netdiffusion}, and FlowChronicle~\cite{cuppers2024flowchronicle}, as well as four general-purpose tabular data generators: E-WGAN-GP~\cite{gulrajani2017ewgan}, CTGAN~\cite{xu2019ctgan}, TVAE~\cite{xu2019ctgan}, and REaLTabFormer~\cite{solatorio2023realtabformer}.\footnote{FlowChronicle only supports CIDDS, and NetDiffusion only PCAP.}
This selection covers a broad range of generative models, including GANs, Diffusion models, Variational Autoencoders, and Transformers.
For comparison, we also include vanilla GPT-2~\cite{radford2019gpt2} and GPT-2 with rejection sampling.
In the latter, the model resamples whenever an inconsistent output is detected, repeating until enough consistent samples are produced.

\mypar{Metrics} 
For each dataset, we generate 10K samples (flow records for CIDDS, packets for Netflix and MAWI, and measurements for MetaDC).
We compare the Jensen–Shannon Divergence (JSD) and Wasserstein/Earth Mover’s Distance (EMD) of the synthetic data generated from different frameworks against the distributions of the original data.
For both metrics, the lower, the better.

\mypar{Fidelity results}
As shown in Fig.~\ref{fig:syngen} (and Fig.~\ref{fig:syngen2} in Appendix~\ref{apdx:syngen}), \sys achieves on average 1.94\x lower JSD and 27.9\x lower normalized EMD across the four datasets (excluding the maximum in each case).
Unlike other frameworks such as NetDiffusion, which perform well on one metric (JSD) but poorly on another (EMD), \sys delivers consistently strong performance on both.

\mypar{Rule-compliance results}
Fig.~\ref{fig:vrank} in Appendix~\ref{apdx:rule-compilance} shows the evaluation of data generators in terms of violations against our benchmark rules.
By enforcing rules during inference, \sys guarantees compliance and achieves zero violations on all datasets except MAWI.
In contrast, other frameworks exhibit high violation rates, exposing serious breaks in network semantics within their generated data.
Critically, network-specific frameworks do not outperform general-purpose models in rule compliance, suggesting that domain knowledge is not well integrated into their designs.
Moreover, as shown in Fig.~\ref{fig:vstats} in Appendix~\ref{apdx:rule-compilance}, the CDF of rule violations reveals that more than 50\% of the rules are violated by multiple generated samples.

\subsection{Network Traffic Forecasting} \label{sssec:forecast}

\mypar{Baselines}
For traffic forecasting, we compare \sys against four traditional regression methods: XGBoost, LightGBM, ARIMA, and ConvLSTM, as well as vanilla Transformer, RealTabFormer~\cite{solatorio2023realtabformer}, vanilla GPT-2~\cite{radford2019gpt2}, and ASTGN (Attention-based Spatial-Temporal Graph Network)~\cite{peng2024astgn}.
Note that ASTGN is specifically designed for time-series network traffic forecasting.

\mypar{Metrics}
We provide all models with network signals (ECN markings, connection count, and ingress/egress retransmissions) of the past five 50ms-intervals and asking them to predict ingress/egress traffic volume of the next 50ms interval.
We evaluate forecasting performance using six metrics: mean absolute error (MAE), root mean squared error (RMSE), normalized scale-free RMSE (NRMSE), symmetric mean absolute percentage error (sMAPE), explained variance (R2), and linear correlation (PearsonR).

\mypar{Results}
Fig.~\ref{fig:forecast} shows the performance of nine forecasting methods ranked from best to worst.
Without enforcing rules, vanilla GPT-2 ranks near the bottom as generic LMs are not inherently good at handling numbers~\cite{mirzadeh2024gsm,ahn2024large,qian2022limitations}. (Full results are reported in Fig.~\ref{fig:forecast2},  Appendix~\ref{apdx:forecast}.)
By enforcing learned network rules, \sys boosts the vanilla GPT-2 model by 15.04\%--42.67\% across different metrics, demonstrating knowledge enforcement that can reduce the need for expensive retraining for every task.
While \sys is not the top performer, it achieves accuracy comparable to the specialized traffic predictor ASTGN~\cite{peng2024astgn}.

\begin{figure}[t]
    \centering
    \begin{adjustbox}{width=1\linewidth,center=0pt}
    \includegraphics[width=\linewidth]{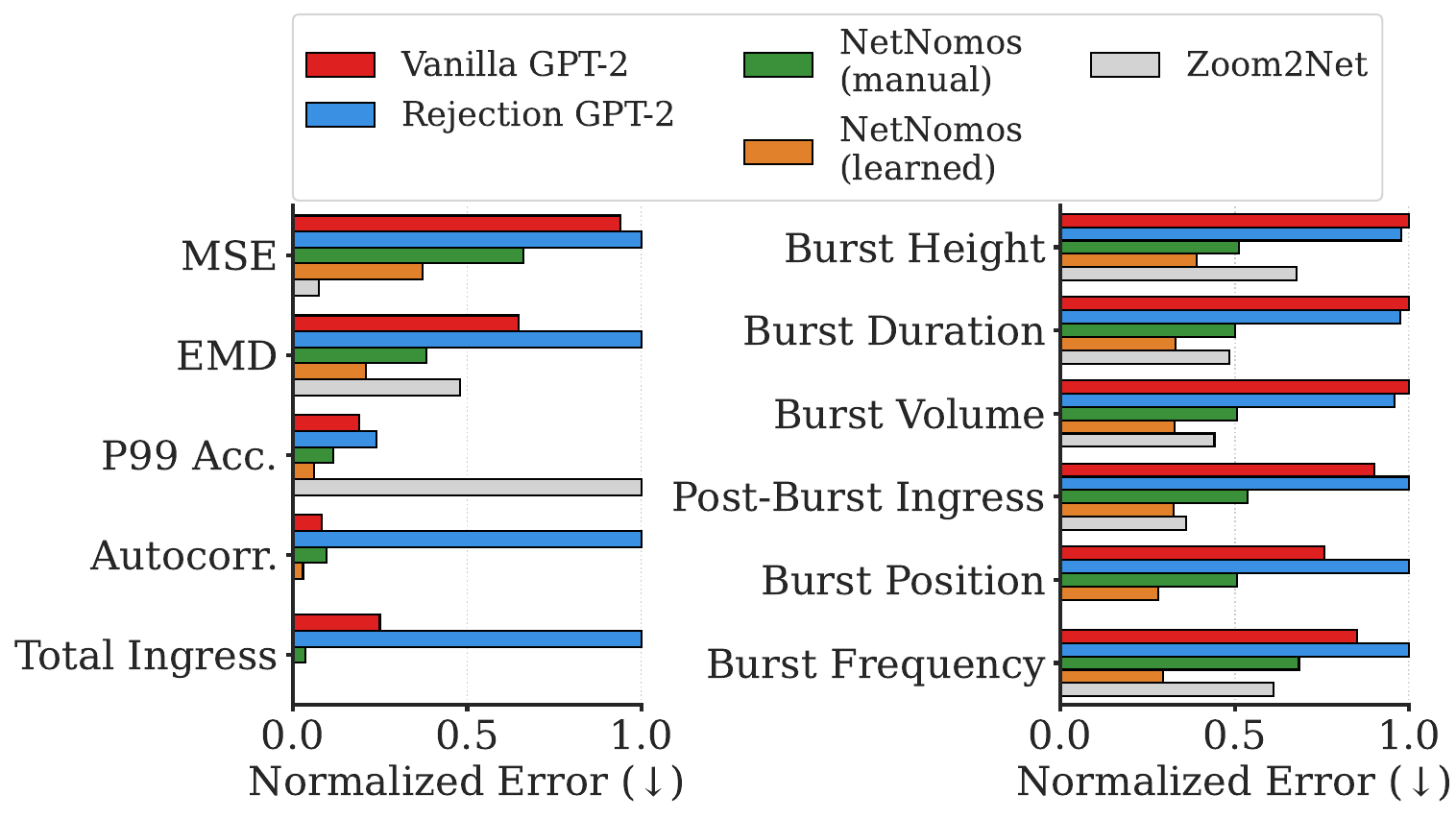}
    \end{adjustbox}
\caption{\sys improves both imputation accuracy (left) and downstream task performance (right) of the generic GPT-2  via logic enforcement, achieving on-par (and often better) results compared to SOTA Zoom2Net~\cite{gong2024zoom2net}.} 
\label{fig:imputation}
\end{figure}

\subsection{Network Telemetry Imputation} \label{sssec:impute}

\mypar{Methodology}
The goal of this use case is to impute 1ms ingress values given aggregated network signals sampled at 50ms intervals.
We compare \sys against three baselines: the SOTA telemetry imputer Zoom2Net~\cite{gong2024zoom2net}, vanilla GPT-2, and GPT-2 with rejection sampling.
For \sys, we evaluate two rule sets: \texttt{manual} and \texttt{learned}.
The \texttt{manual} rules correspond to the four constraints manually identified by the authors of Zoom2Net (C4--C7 in \cite{gong2024zoom2net}), while the \texttt{learned} rules are automatically learned by \sys.

\mypar{Results}
Fig.~\ref{fig:imputation} reports performance both on imputation accuracy and on a downstream task, burst analysis~\cite{ghabashneh2022millisampler}.
Enforcing \texttt{manual} rules improves GPT-2’s accuracy (Fig.~\ref{fig:imputation}, left), but still lags behind Zoom2Net.
Rejection sampling, by contrast, reduces accuracy: it distorts the LM’s distribution by suppressing near-correct outputs and forcing sampling from unrelated regions.
With its \texttt{learned} rules, \sys matches or surpasses Zoom2Net on EMD and p99 accuracy.
When guided by \sys, GPT-2 outperforms Zoom2Net on all downstream tasks except Burst Position for which Zoom2net uses a specially crafted rule.
This demonstrates that automatically learned rules improve the performance of the downstream tasks (not just direct output), even compared to hand-picked rules.
The remaining shortfall on time-sensitive metrics (\eg autocorrelation, Burst Position) likely arises from two factors: GPT-2’s general-purpose architecture and the limited temporal expressiveness of learned rules, constrained by the context window of $\Gamma$.
Developing methods to learn richer temporal constraints is an important direction for future work and could unlock even greater benefits for \sys.

\section{Conclusion}
We present \sys, the first framework that automatically learns rules from large volumes of network data and enforces them during language model generation to produce compliant outputs.
This work takes an important step toward building practical neurosymbolic systems for networking.

\clearpage

\section*{Acknowledgment}

We thank the anonymous reviewers for their helpful feedback.
We also thank Aarti Gupta for her insightful suggestions and guidance during the revision of the manuscript.
We appreciate our shepherd, Suman Nath, for his support and thoughtful feedback.
This work was supported by the National Science Foundation (NSF) through Grants CNS-2442625 and CNS-2319442.






\bibliographystyle{plain}
\bibliography{reference,hotnets24-template}

@ARTICLE{red,
author={Sally Floyd and Van {Jacobson}},
journal={IEEE/ACM Transactions on Networking},
title={Random early detection gateways for congestion avoidance},
year={1993},
volume={1},
number={4},
pages={397-413},
ISSN={},
month={Aug},}

@article{tay2021charformer,
  title={Charformer: Fast character transformers via gradient-based subword tokenization},
  author={Tay, Yi and Tran, Vinh Q and Ruder, Sebastian and Gupta, Jai and Chung, Hyung Won and Bahri, Dara and Qin, Zhen and Baumgartner, Simon and Yu, Cong and Metzler, Donald},
  journal={arXiv preprint arXiv:2106.12672},
  year={2021}
}

@inproceedings{yao2021distai,
  title={$\{$DistAI$\}$:$\{$Data-Driven$\}$ automated invariant learning for distributed protocols},
  author={Yao, Jianan and Tao, Runzhou and Gu, Ronghui and Nieh, Jason and Jana, Suman and Ryan, Gabriel},
  booktitle={15th USENIX symposium on operating systems design and implementation (OSDI 21)},
  pages={405--421},
  year={2021}
}

@inproceedings{yao2022duoai,
  title={$\{$DuoAI$\}$: Fast, automated inference of inductive invariants for verifying distributed protocols},
  author={Yao, Jianan and Tao, Runzhou and Gu, Ronghui and Nieh, Jason},
  booktitle={16th USENIX Symposium on Operating Systems Design and Implementation (OSDI 22)},
  pages={485--501},
  year={2022}
}

@inproceedings{hance2021swiss,
  title={Finding invariants of distributed systems: It's a small (enough) world after all},
  author={Hance, Travis and Heule, Marijn and Martins, Ruben and Parno, Bryan},
  booktitle={18th USENIX symposium on networked systems design and implementation (NSDI 21)},
  pages={115--131},
  year={2021}
}

@article{BeurerKellner2024domino,
  title={Guiding LLMs The Right Way: Fast, Non-Invasive Constrained Generation},
  author={Luca Beurer-Kellner and Marc Fischer and Martin T. Vechev},
  journal={International Conference on Machine Learning (ICML)},
  year={2024},
  volume={abs/2403.06988},
}

@article{raffel2020t5,
  title={Exploring the limits of transfer learning with a unified text-to-text transformer},
  author={Raffel, Colin and Shazeer, Noam and Roberts, Adam and Lee, Katherine and Narang, Sharan and Matena, Michael and Zhou, Yanqi and Li, Wei and Liu, Peter J},
  journal={Journal of machine learning research},
  volume={21},
  number={140},
  pages={1--67},
  year={2020}
}

@article{poesia2022synchromesh,
  title={Synchromesh: Reliable code generation from pre-trained language models},
  author={Poesia, Gabriel and Polozov, Oleksandr and Le, Vu and Tiwari, Ashish and Soares, Gustavo and Meek, Christopher and Gulwani, Sumit},
  journal={arXiv preprint arXiv:2201.11227},
  year={2022}
}

@article{geng2023gcd,
  title={Grammar-constrained decoding for structured NLP tasks without finetuning},
  author={Geng, Saibo and Josifoski, Martin and Peyrard, Maxime and West, Robert},
  journal={arXiv preprint arXiv:2305.13971},
  year={2023}
}

@article{radford2019gpt2,
  title={Language Models are Unsupervised Multitask Learners},
  author={Radford, Alec and Wu, Jeffrey and Child, Rewon and Luan, David and Amodei, Dario and Sutskever, Ilya},
  journal={OpenAI Blog},
  volume={1},
  number={8},
  year={2019},
  note={\url{https://cdn.openai.com/better-language-models/language_models_are_unsupervised_multitask_learners.pdf}}
}

@incollection{tretmans2008model,
  title={Model based testing with labelled transition systems},
  author={Tretmans, Jan},
  booktitle={Formal Methods and Testing: An Outcome of the FORTEST Network, Revised Selected Papers},
  pages={1--38},
  year={2008},
  publisher={Springer}
}

@article{classen2012featured,
  title={Featured transition systems: Foundations for verifying variability-intensive systems and their application to LTL model checking},
  author={Classen, Andreas and Cordy, Maxime and Schobbens, Pierre-Yves and Heymans, Patrick and Legay, Axel and Raskin, Jean-Fran{\c{c}}ois},
  journal={IEEE Transactions on Software Engineering},
  volume={39},
  number={8},
  pages={1069--1089},
  year={2012},
  publisher={IEEE}
}

@article{van1994logic,
  title={Logic of transition systems},
  author={Van Benthem, Johan and Bergstra, Jan},
  journal={Journal of Logic, Language and Information},
  volume={3},
  pages={247--283},
  year={1994},
  publisher={Springer}
}

@incollection{ramsey1987epr,
  title={On a problem of formal logic},
  author={Ramsey, Frank P},
  booktitle={Classic Papers in Combinatorics},
  pages={1--24},
  year={1987},
  publisher={Springer}
}

@String{Computing = "Computing" }

@String{Computer = "{IEEE} Computer" }

@String{Academic = "Academic Press" }

@String{Springer = "Springer-Verlag" }

@article{lin2019generating,
  title={Generating high-fidelity, synthetic time series datasets with doppelganger},
  author={Lin, Zinan and Jain, Alankar and Wang, Chen and Fanti, Giulia and Sekar, Vyas},
  journal={arXiv preprint arXiv:1909.13403},
  year={2019}
}

@inproceedings{vaswani2017attention,
 author = {Vaswani, Ashish and Shazeer, Noam and Parmar, Niki and Uszkoreit, Jakob and Jones, Llion and Gomez, Aidan N and Kaiser, \L ukasz and Polosukhin, Illia},
 booktitle = {Advances in Neural Information Processing Systems},
 editor = {I. Guyon and U. Von Luxburg and S. Bengio and H. Wallach and R. Fergus and S. Vishwanathan and R. Garnett},
 pages = {},
 publisher = {Curran Associates, Inc.},
 title = {{Attention is All you Need}},
 url = {https://proceedings.neurips.cc/paper_files/paper/2017/file/3f5ee243547dee91fbd053c1c4a845aa-Paper.pdf},
 volume = {30},
 year = {2017}
}

@inproceedings{ferraiuolo2018hyperflow,
  title={HyperFlow: A processor architecture for nonmalleable, timing-safe information flow security},
  author={Ferraiuolo, Andrew and Zhao, Mark and Myers, Andrew C and Suh, G Edward},
  booktitle={Proceedings of the 2018 ACM SIGSAC Conference on Computer and Communications Security},
  pages={1583--1600},
  year={2018}
}

@inproceedings{jin2024pants,
  title={Robustifying ML-powered Network Classifiers with PANTS},
  author={Minhao Jin and Maria Apostolaki},
  booktitle={34th USENIX Security Symposium (USENIX Security 25)},
  year={2025}
}

@inproceedings{ghabashneh2022millisampler,
  title={A microscopic view of bursts, buffer contention, and loss in data centers},
  author={Ghabashneh, Ehab and Zhao, Yimeng and Lumezanu, Cristian and Spring, Neil and Sundaresan, Srikanth and Rao, Sanjay},
  booktitle={Proceedings of the 22nd ACM Internet Measurement Conference},
  pages={567--580},
  year={2022}
}

@inproceedings{gong2024zoom2net,
  title={Zoom2net: Constrained network telemetry imputation},
  author={Gong, Fengchen and Raghunathan, Divya and Gupta, Aarti and Apostolaki, Maria},
  booktitle={Proceedings of the ACM SIGCOMM 2024 Conference},
  pages={764--777},
  year={2024}
}

@inproceedings{yin2022netshare,
  title={Practical gan-based synthetic ip header trace generation using netshare},
  author={Yin, Yucheng and Lin, Zinan and Jin, Minhao and Fanti, Giulia and Sekar, Vyas},
  booktitle={Proceedings of the ACM SIGCOMM 2022 Conference},
  pages={458--472},
  year={2022}
}

@article{ring2017cidds,
title={ Creation of Flow-Based Data Sets for Intrusion Detection},
author={Ring, Markus and Wunderlich, Sarah and Grüdl, Dominik and Landes, Dieter and Hotho, Andreas},
journal={Journal of Information Warfare},
volume={16},
issue={4},
year={2017},
pages={40–53},
publisher={JIW}}

@article{cuppers2024flowchronicle,
  title={FlowChronicle: Synthetic Network Flow Generation through Pattern Set Mining},
  author={C{\"u}ppers, Joscha and Schoen, Adrien and Blanc, Gregory and Gimenez, Pierre-Francois},
  journal={Proceedings of the ACM on Networking},
  volume={2},
  number={CoNEXT4},
  pages={1--20},
  year={2024},
  publisher={ACM New York, NY, USA}
}

@article{angluin1987lstar,
  title={Learning regular sets from queries and counterexamples},
  author={Angluin, Dana},
  journal={Information and computation},
  volume={75},
  number={2},
  pages={87--106},
  year={1987},
  publisher={Elsevier}
}

@article{fitch1963logical,
  title={A logical analysis of some value concepts1},
  author={Fitch, Frederic B},
  journal={The journal of symbolic logic},
  volume={28},
  number={2},
  pages={135--142},
  year={1963},
  publisher={Cambridge University Press}
}

@incollection{cohen2006langcomplexity,
  title={The complexity of constraint languages},
  author={Cohen, David and Jeavons, Peter},
  booktitle={Foundations of Artificial Intelligence},
  volume={2},
  pages={245--280},
  year={2006},
  publisher={Elsevier}
}

@article{jiang2024netdiffusion,
  title={Netdiffusion: Network data augmentation through protocol-constrained traffic generation},
  author={Jiang, Xi and Liu, Shinan and Gember-Jacobson, Aaron and Bhagoji, Arjun Nitin and Schmitt, Paul and Bronzino, Francesco and Feamster, Nick},
  journal={Proceedings of the ACM on Measurement and Analysis of Computing Systems},
  volume={8},
  number={1},
  pages={1--32},
  year={2024},
  publisher={ACM New York, NY, USA}
}

@article{bronzino2019netflix,
  title={Inferring streaming video quality from encrypted traffic: Practical models and deployment experience},
  author={Bronzino, Francesco and Schmitt, Paul and Ayoubi, Sara and Martins, Guilherme and Teixeira, Renata and Feamster, Nick},
  journal={Proceedings of the ACM on Measurement and Analysis of Computing Systems},
  volume={3},
  number={3},
  pages={1--25},
  year={2019},
  publisher={ACM New York, NY, USA}
}

@inproceedings{schoen2024rulecidds,
  title={A Tale of Two Methods: Unveiling the limitations of GAN and the Rise of Bayesian Networks for Synthetic Network Traffic Generation},
  author={Schoen, Adrien and Blanc, Gregory and Gimenez, Pierre-Fran{\c{c}}ois and Han, Yufei and Majorczyk, Fr{\'e}d{\'e}ric and Me, Ludovic},
  booktitle={2024 IEEE European Symposium on Security and Privacy Workshops (EuroS\&PW)},
  pages={273--286},
  year={2024},
  organization={IEEE}
}

@article{ring2019rulecidds,
  title={Flow-based network traffic generation using generative adversarial networks},
  author={Ring, Markus and Schl{\"o}r, Daniel and Landes, Dieter and Hotho, Andreas},
  journal={Computers \& Security},
  volume={82},
  pages={156--172},
  year={2019},
  publisher={Elsevier}
}

@article{xu2019ctgan,
  title={Modeling tabular data using conditional gan},
  author={Xu, Lei and Skoularidou, Maria and Cuesta-Infante, Alfredo and Veeramachaneni, Kalyan},
  journal={Advances in neural information processing systems},
  volume={32},
  year={2019}
}

@article{solatorio2023realtabformer,
  title={REaLTabFormer: Generating Realistic Relational and Tabular Data using Transformers},
  author={Solatorio, Aivin V. and Dupriez, Olivier},
  journal={arXiv preprint arXiv:2302.02041},
  year={2023}
}

@misc{rfc793,
  author       = {Jon Postel},
  title        = {Transmission Control Protocol},
  howpublished = {RFC 793},
  series       = {Request for Comments},
  number       = {793},
  year         = {1981},
  month        = sep,
  publisher    = {RFC Editor},
  doi          = {10.17487/RFC0793},
  url          = {https://www.rfc-editor.org/info/rfc793}
}

@inproceedings{dershowitz2006scalable,
  title={A scalable algorithm for minimal unsatisfiable core extraction},
  author={Dershowitz, Nachum and Hanna, Ziyad and Nadel, Alexander},
  booktitle={Theory and Applications of Satisfiability Testing-SAT 2006: 9th International Conference, Seattle, WA, USA, August 12-15, 2006. Proceedings 9},
  pages={36--41},
  year={2006},
  organization={Springer}
}

@inproceedings{ignatiev2015smallest,
  title={Smallest MUS extraction with minimal hitting set dualization},
  author={Ignatiev, Alexey and Previti, Alessandro and Liffiton, Mark and Marques-Silva, Joao},
  booktitle={International Conference on Principles and Practice of Constraint Programming},
  pages={173--182},
  year={2015},
  organization={Springer}
}

@inproceedings{nadel2010boosting,
  title={Boosting minimal unsatisfiable core extraction},
  author={Nadel, Alexander},
  booktitle={Formal methods in computer aided design},
  pages={221--229},
  year={2010},
  organization={IEEE}
}

@article{gulrajani2017ewgan,
  title={Improved training of wasserstein gans},
  author={Gulrajani, Ishaan and Ahmed, Faruk and Arjovsky, Martin and Dumoulin, Vincent and Courville, Aaron C},
  journal={Advances in neural information processing systems},
  volume={30},
  year={2017}
}

@article{pei2007hmine,
  title={H-Mine: Fast and space-preserving frequent pattern mining in large databases},
  author={Pei, Jian and Han, Jiawei and Lu, Hongjun and Nishio, Shojiro and Tang, Shiwei and Yang, Dongqing},
  journal={IIE transactions},
  volume={39},
  number={6},
  pages={593--605},
  year={2007},
  publisher={Taylor \& Francis}
}

@article{chu2013fastdc,
  title={Discovering denial constraints},
  author={Chu, Xu and Ilyas, Ihab F and Papotti, Paolo},
  journal={Proceedings of the VLDB Endowment},
  volume={6},
  number={13},
  pages={1498--1509},
  year={2013},
  publisher={VLDB Endowment}
}

@article{pena2022fastdc++,
  title={Fast algorithms for denial constraint discovery},
  author={Pena, Eduardo HM and Porto, Fabio and Naumann, Felix},
  journal={Proceedings of the VLDB Endowment},
  volume={16},
  number={4},
  pages={684--696},
  year={2022},
  publisher={VLDB Endowment}
}

@misc{rfc768,
  author       = {Jon Postel},
  title        = {User Datagram Protocol},
  howpublished = {RFC 768},
  year         = {1980},
  month        = aug,
  institution  = {Internet Engineering Task Force},
  url          = {https://www.rfc-editor.org/rfc/rfc768},
}

@misc{rfc791,
  author       = {Jon Postel},
  title        = {Internet Protocol},
  howpublished = {RFC 791},
  year         = {1981},
  month        = sep,
  institution  = {Internet Engineering Task Force},
  url          = {https://www.rfc-editor.org/rfc/rfc791},
}

@misc{rfc1122,
  author       = {R. Braden},
  title        = {Requirements for Internet Hosts -- Communication Layers},
  howpublished = {RFC 1122},
  year         = {1989},
  month        = oct,
  institution  = {Internet Engineering Task Force},
  url          = {https://www.rfc-editor.org/rfc/rfc1122},
}

@misc{rfc4443,
  author       = {A. Conta and S. Deering and M. Gupta},
  title        = {Internet Control Message Protocol (ICMPv6) for the Internet Protocol Version 6 (IPv6) Specification},
  howpublished = {RFC 4443},
  year         = {2006},
  month        = mar,
  institution  = {Internet Engineering Task Force},
  url          = {https://www.rfc-editor.org/rfc/rfc4443},
}

@misc{rfc6335,
  author       = {M. Cotton and L. Eggert and J. Touch and M. Westerlund and S. Cheshire},
  title        = {Internet Assigned Numbers Authority (IANA) Procedures for the Management of the Service Name and Transport Protocol Port Number Registry},
  howpublished = {RFC 6335},
  year         = {2011},
  month        = aug,
  institution  = {Internet Engineering Task Force},
  url          = {https://www.rfc-editor.org/rfc/rfc6335},
}

@misc{rfc7605,
  author       = {J. Touch},
  title        = {Recommendations on Using Assigned Transport Port Numbers},
  howpublished = {RFC 7605},
  year         = {2015},
  month        = aug,
  institution  = {Internet Engineering Task Force},
  url          = {https://www.rfc-editor.org/rfc/rfc7605},
}

@misc{rfc8200,
  author       = {S. Deering and R. Hinden},
  title        = {Internet Protocol, Version 6 (IPv6) Specification},
  howpublished = {RFC 8200},
  year         = {2017},
  month        = jul,
  institution  = {Internet Engineering Task Force},
  url          = {https://www.rfc-editor.org/rfc/rfc8200},
}

@misc{rfc9293,
  author       = {W. Eddy, Ed.},
  title        = {Transmission Control Protocol (TCP)},
  howpublished = {RFC 9293},
  year         = {2022},
  month        = aug,
  institution  = {Internet Engineering Task Force},
  url          = {https://www.rfc-editor.org/rfc/rfc9293},
}

@misc{rfc9673,
  author       = {R. Hinden and G. Fairhurst},
  title        = {IPv6 Hop-by-Hop Options Processing Procedures},
  howpublished = {RFC 9673},
  year         = {2024},
  month        = oct,
  institution  = {Internet Engineering Task Force},
  url          = {https://www.rfc-editor.org/rfc/rfc9673},
}

@inproceedings{mawi2006traffic,
  title     = {The MAWI Working Group Traffic Archive},
  author    = {Kenji Cho and Koushirou Mitsuya and Akira Kato},
  booktitle = {Proceedings of the 2000 USENIX Annual Technical Conference (FREENIX Track)},
  year      = {2000},
  pages     = {263--270},
  url       = {http://mawi.wide.ad.jp/mawi/}
}

@inproceedings{ma2017netflix,
  title     = {Neural Networks for Modeling Netflix’s Dynamic Pricing System},
  author    = {Ma, Qian and Bharambe, Ashwin and Harchol-Balter, Mor and Scheller-Wolf, Alan},
  booktitle = {Proceedings of the 2017 ACM SIGMETRICS International Conference on Measurement and Modeling of Computer Systems},
  year      = {2017},
  pages     = {1--13},
  doi       = {10.1145/3078505.3078532}
}

@misc{iana-protocol-numbers,
  author       = {{Internet Assigned Numbers Authority}},
  title        = {Assigned Internet Protocol Numbers},
  howpublished = {\url{https://www.iana.org/assignments/protocol-numbers/}},
  note         = {Last Updated: 2025-07-11},
  year         = {2025}
}

@misc{iana-service-names,
  author       = {{Internet Assigned Numbers Authority}},
  title        = {Service Name and Transport Protocol Port Number Registry},
  howpublished = {\url{https://www.iana.org/assignments/service-names-port-numbers/}},
  note         = {Last Updated: 2025-08-26},
  year         = {2025}
}

@inproceedings{jacobs2022trustee,
  title     = {{AI/ML for Network Security: The Emperor has no Clothes}},
  author    = {Arthur S. Jacobs and Roman Beltiukov and Walter Willinger and Ronaldo A. Ferreira and Arpit Gupta and Lisandro Z. Granville},
  booktitle = {Proceedings of the 2022 ACM SIGSAC Conference on Computer and Communications Security (CCS '22)},
  year      = {2022},
  publisher = {ACM},
  address   = {Los Angeles, CA, USA},
  doi       = {10.1145/3548606.3560609},
  url       = {https://trusteeml.github.io/}
}

@misc{rfc7323,
  author       = {E. Borman and B. Braden and V. Jacobson and R. Scheffenegger},
  title        = {TCP Extensions for High Performance},
  howpublished = {RFC 7323},
  year         = {2014},
  month        = sep,
  publisher    = {Internet Engineering Task Force},
  url          = {https://www.rfc-editor.org/rfc/rfc7323},
  doi          = {10.17487/RFC7323}
}

@article{chu2025netssm,
  title={NetSSM: Multi-Flow and State-Aware Network Trace Generation using State-Space Models},
  author={Chu, Andrew and Jiang, Xi and Liu, Shinan and Bhagoji, Arjun and Bronzino, Francesco and Schmitt, Paul and Feamster, Nick},
  journal={arXiv preprint arXiv:2503.22663},
  year={2025}
}

@article{reiter1987minhitset,
  title={A theory of diagnosis from first principles},
  author={Reiter, Raymond},
  journal={Artificial intelligence},
  volume={32},
  number={1},
  pages={57--95},
  year={1987},
  publisher={Elsevier}
}

@inproceedings{zhang2025basilisk,
  title={Basilisk: Using Provenance Invariants to Automate Proofs of Undecidable Protocols},
  author={Zhang, Tony Nuda and Singh, Keshav and Chajed, Tej and Kapritsos, Manos and Parno, Bryan},
  booktitle={19th USENIX Symposium on Operating Systems Design and Implementation (OSDI 25)},
  pages={1--17},
  year={2025}
}

@inproceedings{padon2016ivy,
  title={Ivy: safety verification by interactive generalization},
  author={Padon, Oded and McMillan, Kenneth L and Panda, Aurojit and Sagiv, Mooly and Shoham, Sharon},
  booktitle={Proceedings of the 37th ACM SIGPLAN Conference on Programming Language Design and Implementation},
  pages={614--630},
  year={2016}
}

@inproceedings{ma2019i4,
  title={I4: incremental inference of inductive invariants for verification of distributed protocols},
  author={Ma, Haojun and Goel, Aman and Jeannin, Jean-Baptiste and Kapritsos, Manos and Kasikci, Baris and Sakallah, Karem A},
  booktitle={Proceedings of the 27th ACM Symposium on Operating Systems Principles},
  pages={370--384},
  year={2019}
}

@inproceedings{koenig2020folic3,
  title={First-order quantified separators},
  author={Koenig, Jason R and Padon, Oded and Immerman, Neil and Aiken, Alex},
  booktitle={Proceedings of the 41st ACM SIGPLAN conference on programming language design and implementation},
  pages={703--717},
  year={2020}
}

@inproceedings{hsieh2024netvigil,
  title={$\{$NetVigil$\}$: Robust and $\{$Low-Cost$\}$ Anomaly Detection for $\{$East-West$\}$ Data Center Security},
  author={Hsieh, Kevin and Wong, Mike and Segarra, Santiago and Mani, Sathiya Kumaran and Eberl, Trevor and Panasyuk, Anatoliy and Netravali, Ravi and Chandra, Ranveer and Kandula, Srikanth},
  booktitle={21st USENIX Symposium on Networked Systems Design and Implementation (NSDI 24)},
  pages={1771--1789},
  year={2024}
}

@inproceedings{li2024reasoning,
  title={Reasoning about network traffic load property at production scale},
  author={Li, Ruihan and Ye, Fangdan and Yuan, Yifei and Yang, Ruizhen and Tian, Bingchuan and Guo, Tianchen and Wu, Hao and Zhu, Xiaobo and Guan, Zhongyu and Ma, Qing and others},
  booktitle={21st USENIX Symposium on Networked Systems Design and Implementation (NSDI 24)},
  pages={1063--1082},
  year={2024}
}

@inproceedings{zhang2022differential,
  title={Differential network analysis},
  author={Zhang, Peng and Gember-Jacobson, Aaron and Zuo, Yueshang and Huang, Yuhao and Liu, Xu and Li, Hao},
  booktitle={19th USENIX Symposium on Networked Systems Design and Implementation (NSDI 22)},
  pages={601--615},
  year={2022}
}

@inproceedings{gember2015management,
  title={Management plane analytics},
  author={Gember-Jacobson, Aaron and Wu, Wenfei and Li, Xiujun and Akella, Aditya and Mahajan, Ratul},
  booktitle={Proceedings of the 2015 Internet Measurement Conference},
  pages={395--408},
  year={2015}
}

@inproceedings{yu2013software,
  title={Software $\{$Defined$\}$$\{$Traffic$\}$ Measurement with $\{$OpenSketch$\}$},
  author={Yu, Minlan and Jose, Lavanya and Miao, Rui},
  booktitle={10th USENIX symposium on networked systems design and implementation (NSDI 13)},
  pages={29--42},
  year={2013}
}

@article{stupan2022niaarm,
  title={Niaarm: a minimalistic framework for numerical association rule mining},
  author={Stupan, {\v{Z}}iga and Fister, Iztok},
  journal={Journal of Open Source Software},
  volume={7},
  number={77},
  pages={4448},
  year={2022}
}

@book{domingos2015master,
  title={The master algorithm: How the quest for the ultimate learning machine will remake our world},
  author={Domingos, Pedro},
  year={2015},
  publisher={Basic Books}
}

@inproceedings{lewis2000cart,
  title={An introduction to classification and regression tree (CART) analysis},
  author={Lewis, Roger J and others},
  booktitle={Annual meeting of the society for academic emergency medicine in San Francisco, California},
  volume={14},
  pages={106},
  year={2000},
  organization={Department of Emergency Medicine Harbor-UCLA Medical Center Torrance San~…}
}

@inproceedings{winterer2020semanticfusion,
  title={Validating SMT solvers via semantic fusion},
  author={Winterer, Dominik and Zhang, Chengyu and Su, Zhendong},
  booktitle={Proceedings of the 41st ACM SIGPLAN Conference on programming language design and implementation},
  pages={718--730},
  year={2020}
}

@inproceedings{ben2024cafa,
  title={CaFA: Cost-aware, Feasible Attacks With Database Constraints Against Neural Tabular Classifiers},
  author={Ben-Tov, Matan and Deutch, Daniel and Frost, Nave and Sharif, Mahmood},
  booktitle={2024 IEEE Symposium on Security and Privacy (SP)},
  pages={1345--1364},
  year={2024},
  organization={IEEE}
}

@article{peng2024astgn,
  title={Network traffic prediction with attention-based spatial--temporal graph network},
  author={Peng, Yufei and Guo, Yingya and Hao, Run and Xu, Chengzhe},
  journal={Computer Networks},
  volume={243},
  pages={110296},
  year={2024},
  publisher={Elsevier}
}

@article{sivaroopan2025comprehensive,
  title={A Comprehensive Survey on Network Traffic Synthesis: From Statistical Models to Deep Learning},
  author={Sivaroopan, Nirhoshan and Silva, Kaushitha and Madarasingha, Chamara and Dahanayaka, Thilini and Jourjon, Guillaume and Jayasumana, Anura and Thilakarathna, Kanchana},
  journal={arXiv preprint arXiv:2507.01976},
  year={2025}
}

@article{guthula2023netfound,
  title={netFound: Foundation model for network security},
  author={Guthula, Satyandra and Beltiukov, Roman and Battula, Navya and Guo, Wenbo and Gupta, Arpit},
  journal={arXiv preprint arXiv:2310.17025},
  year={2023}
}

@article{Basin2025Gap,
author = {Basin, David and Foster, Nate and McMillan, Kenneth L. and Namjoshi, Kedar S. and Nita-Rotaru, Cristina and Smith, Jonathan M. and Zave, Pamela and Zuck, Lenore D.},
title = {It Takes a Village: Bridging the Gaps between Current and Formal Specifications for Protocols},
year = {2025},
issue_date = {August 2025},
publisher = {Association for Computing Machinery},
address = {New York, NY, USA},
volume = {68},
number = {8},
issn = {0001-0782},
url = {https://doi.org/10.1145/3706572},
doi = {10.1145/3706572},
journal = {Commun. ACM},
month = jul,
pages = {50–61},
numpages = {12}
}

@article{zhou2024rulearena,
  title={Rulearena: A benchmark for rule-guided reasoning with llms in real-world scenarios},
  author={Zhou, Ruiwen and Hua, Wenyue and Pan, Liangming and Cheng, Sitao and Wu, Xiaobao and Yu, En and Wang, William Yang},
  journal={arXiv preprint arXiv:2412.08972},
  year={2024}
}

@article{tam2024let,
  title={Let me speak freely? a study on the impact of format restrictions on performance of large language models},
  author={Tam, Zhi Rui and Wu, Cheng-Kuang and Tsai, Yi-Lin and Lin, Chieh-Yen and Lee, Hung-yi and Chen, Yun-Nung},
  journal={arXiv preprint arXiv:2408.02442},
  year={2024}
}

@article{hartmanis1982computers,
  title={Computers and intractability: a guide to the theory of np-completeness (michael r. garey and david s. johnson)},
  author={Hartmanis, Juris},
  journal={Siam Review},
  volume={24},
  number={1},
  pages={90},
  year={1982},
  publisher={Society for Industrial and Applied Mathematics}
}

@article{mirzadeh2024gsm,
  title={Gsm-symbolic: Understanding the limitations of mathematical reasoning in large language models},
  author={Mirzadeh, Iman and Alizadeh, Keivan and Shahrokhi, Hooman and Tuzel, Oncel and Bengio, Samy and Farajtabar, Mehrdad},
  journal={arXiv preprint arXiv:2410.05229},
  year={2024}
}

@article{ahn2024large,
  title={Large language models for mathematical reasoning: Progresses and challenges},
  author={Ahn, Janice and Verma, Rishu and Lou, Renze and Liu, Di and Zhang, Rui and Yin, Wenpeng},
  journal={arXiv preprint arXiv:2402.00157},
  year={2024}
}

@article{qian2022limitations,
  title={Limitations of language models in arithmetic and symbolic induction},
  author={Qian, Jing and Wang, Hong and Li, Zekun and Li, Shiyang and Yan, Xifeng},
  journal={arXiv preprint arXiv:2208.05051},
  year={2022}
}

@article{papenbrock2015functional,
  title={Functional dependency discovery: An experimental evaluation of seven algorithms},
  author={Papenbrock, Thorsten and Ehrlich, Jens and Marten, Jannik and Neubert, Tommy and Rudolph, Jan-Peer and Sch{\"o}nberg, Martin and Zwiener, Jakob and Naumann, Felix},
  journal={Proceedings of the VLDB Endowment},
  volume={8},
  number={10},
  pages={1082--1093},
  year={2015},
  publisher={VLDB Endowment}
}

@inproceedings{papenbrock2016hybrid,
  title={A hybrid approach to functional dependency discovery},
  author={Papenbrock, Thorsten and Naumann, Felix},
  booktitle={proceedings of the 2016 International Conference on Management of Data},
  pages={821--833},
  year={2016}
}

@book{zhang2002association,
  title={Association rule mining: models and algorithms},
  author={Zhang, Chengqi and Zhang, Shichao},
  year={2002},
  publisher={Springer}
}

@inproceedings{wang2002top,
  title={Top down fp-growth for association rule mining},
  author={Wang, Ke and Tang, Liu and Han, Jiawei and Liu, Junqiang},
  booktitle={Pacific-Asia conference on knowledge discovery and data mining},
  pages={334--340},
  year={2002},
  organization={Springer}
}

@book{boolos2007computability,
  title     = {Computability and Logic},
  author    = {Boolos, George S. and Burgess, John P. and Jeffrey, Richard C.},
  edition   = {5},
  year      = {2007},
  publisher = {Cambridge University Press},
  address   = {Cambridge, UK}
}

@inproceedings{he2025lejit,
  author    = {Hongyu Hè and Maria Apostolaki},
  title     = {Just-in-Time Logic Enforcement: A new paradigm of combining statistical and symbolic reasoning for network management},
  booktitle = {Proceedings of the 24th ACM Workshop on Hot Topics in Networks (HotNets)},
  year      = {2025},
}

@book{bradley2007calculus,
  title={The calculus of computation: decision procedures with applications to verification},
  author={Bradley, Aaron R and Manna, Zohar},
  year={2007},
  publisher={Springer}
}

@article{jin2025tracebleed,
  title={Assessing User Privacy Leakage in Synthetic Packet Traces: An Attack-Grounded Approach},
  author={Jin, Minhao and He, Hongyu and Apostolaki, Maria},
  journal={arXiv preprint arXiv:2508.11742},
  year={2025}
}

@misc{netnomos,
  title = {NetNomos code repository},
  author = {Hongyu H\`e},
  howpublished = {\url{https://github.com/HongyuHe/NetNomos}},
}

\appendix

\section{Proof for Theorem~\ref{thm:hitset}} \label{apdx:proof1}

We construct a proof using a modified formulation of the reduction from clause learning to finding minimal set covers~\cite{chu2013fastdc}.
Let $\mathcal{C}$ denote the set of propositional clauses derived from the grammar $\Gamma$.  
For each clause $c \in \mathcal{C}$, define its \emph{evidence set}
\[
E_c := \{ e \in D \mid e \models c \}.
\]
A constraint of disjuncts is a finite set of clauses $C \subseteq \mathcal{C}$ interpreted as
$\bigvee_{c \in C} c$. By construction,
\[
C \text{ is consistent with } D 
\;\;\Longleftrightarrow\;\;
\bigcup_{c \in C} E_c = D.
\]

\begin{itemize}[leftmargin=*]
    \item[$\Longrightarrow$] Suppose $C \subseteq \mathcal{C}$ is consistent. Then by the equivalence above,
$\bigcup_{c \in C} E_c = D$. Hence $C$ itself is a hitting set of clauses that covers $D$.

    \item[$\Longleftarrow$] Conversely, suppose $H \subseteq \mathcal{C}$ is a hitting set:  
$\bigcup_{c \in H} E_c = D$. Consider the constraint $C_H = \bigvee_{c \in H} c$.  
For any $e \in D$, since $e \in \bigcup_{c \in H} E_c$, there exists $c \in H$ with $e \models c$.  
Thus $e \models C_H$, so $C_H$ is consistent with $D$.
\end{itemize}
\noindent Therefore, learning a consistent constraint is equivalent to finding a hitting set of clauses covering $D$. 
\hfill $\qed$

\section{Network Dataset and Corresponding Benchmark Rulesets}
\label{apdx:ds_and_rules}

Table~\ref{tab:datasets} describes the datasets used for evaluation and the corresponding benchmark rulesets.

\begin{table}[t]
    \centering
    \begin{adjustbox}{width=1.\linewidth,center}
    \begin{tabular}{@{} llp{0.34\linewidth}p{0.5\linewidth} @{}}
    \toprule
    \textbf{Dataset} & \textbf{Format} & \textbf{Benchmark Ruleset} & \textbf{Description} \\
    \midrule
    CIDDS~\cite{ring2017cidds,ring2019rulecidds,schoen2024rulecidds}   
    & NetFlow & 
        Total rules: 72
      \begin{itemize}[leftmargin=*,nosep]
        \item Deployment: 30
        \item Protocol: 42
        \item Principle: 0
      \end{itemize}
    & Normal and attack traffic collected from an enterprise network. \\
    \midrule  
    Netflix~\cite{bronzino2019netflix}   
    & PCAP & 
        Total rules: 33
      \begin{itemize}[leftmargin=*,nosep]
        \item Deployment: 0
        \item Protocol: 33
        \item Principle: 0
      \end{itemize}
    & Video streaming traffic with typical client-server interactions. \\
    \midrule
    MAWI~\cite{mawi2006traffic}   
    & PCAP & 
        Total rules: 32
      \begin{itemize}[leftmargin=*,nosep]
        \item Deployment: 0
        \item Protocol: 32
        \item Principle: 0
      \end{itemize}
    & Backbone Internet traffic traces collected at a trans-Pacific link. \\
    \midrule
    MetaDC~\cite{ghabashneh2022millisampler}   
    & Datacenter logs & 
        Total rules: 10
      \begin{itemize}[leftmargin=*,nosep]
        \item Deployment: 0
        \item Protocol: 0
        \item Principle: 10
      \end{itemize}
    & Millisecond-scale traffic traces from Meta’s datacenters, capturing burstiness, buffer contention, and packet loss. \\
    \bottomrule
    \end{tabular}
    \end{adjustbox}
    \caption{Datasets used together with statistics of their corresponding benchmark rulesets for evaluation. \sys generalizes across diverse real-world network datasets.}
    \label{tab:datasets}
\end{table}

\section{Samples of Benchmark Rules}
\label{apdx:expl-rules}

Tables~\ref{tab:pcap_bench}--\ref{tab:meta_bench} provide samples of the network rules and their corresponding semantics.

\section{Evaluation Rulesets for Evaluating Semantic Filtering}
\label{apdx:filter_rulesets}

Table~\ref{tab:filter_rulesets} summarizes the rulesets used for evaluating semantic filtering.

\begin{table}[!h]
  \centering
 \begin{adjustbox}{width=0.75\linewidth,center=0pt}    
  \renewcommand{\arraystretch}{1}
  \begin{tabular}{lrr}
    \toprule
    \textbf{Dataset} & \textbf{Meaningful Rules} & \textbf{Meaningless Rules} \\
    \midrule
    CIDDS   & 72 & 176 \\
    Netflix & 33 & 82  \\
    MAWI    & 35 & 24  \\
    MetaDC  & 10 & 18  \\
    \bottomrule
  \end{tabular}
  \end{adjustbox}
  \caption{Rulesets used for evaluating semantic rule-filtering.}
  \label{tab:filter_rulesets}
\end{table}

\section{Baselines for Rule Learning}
\label{apdx:baselines}

We compare \sys to five rule-learning methods from three domains, namely, logic programming (LP), database (DB), and machine learning (ML):
\begin{itemize}[leftmargin=*]
\item (LP) DuoAI~\cite{yao2022duoai}: a SOTA method for learning invariants of distributed protocols.  
It constructs a minimal implication graph and enumerates the formula space, starting from the strongest and weakening iteratively.  
Because we do not assume the existence of formal models of the networks of the corresponding datasets, we omit DuoAI’s post-learning refinement with IVy~\cite{padon2016ivy} and only compare against its method for learning candidate invariants.  

\item (DB) FastDC~\cite{chu2013fastdc,pena2022fastdc++}: a SOTA method for discovering denial constraints (DCs) in databases.
DCs are the \textit{most expressive} integrity constraints. We negate learned DCs to derive positive rules, since a DC specifies what must not happen.

\item (DB) H-Mine~\cite{pei2007hmine}: a SOTA association rule mining method that learns ``if-then'' implication patterns between variables in large datasets.  
To apply it to network data, we encode records as boolean variable-value transactions.  
It does not support numerical variables.\footnote{Recent work on numerical association rule learning~\cite{stupan2022niaarm} did not yield meaningful rules, so we omit its results.}  

\item (ML) FlowChronicle~\cite{cuppers2024flowchronicle}: a SOTA network data generator that employs sequential pattern mining and defines a constraint language for network rules.  
Its implementation is tailored to the CIDDS dataset, and adapting it to other datasets would require substantial changes.  
Thus, we only evaluate it on CIDDS.  

\item (ML) Decision tree (DT): Decision paths in DTs correspond to linear combinations of propositions~\cite{domingos2015master}.  
We train a CART-style DT~\cite{lewis2000cart} on each dataset and extract their decision paths as logical implications.  

\item (ML) DTs from GANs: We train DTs as student models on the decisions made by discriminators of NetShare~\cite{yin2022netshare} and CTGAN~\cite{xu2019ctgan}.  
We then extract decision paths from these student models as learned rules.  

\end{itemize}

\section{Expressiveness of Rule Learning}
\label{apdx:fig-express}

\begin{figure}[t]
    \centering
    \begin{adjustbox}{width=1\linewidth,center=0pt}
    \includegraphics[width=\linewidth]{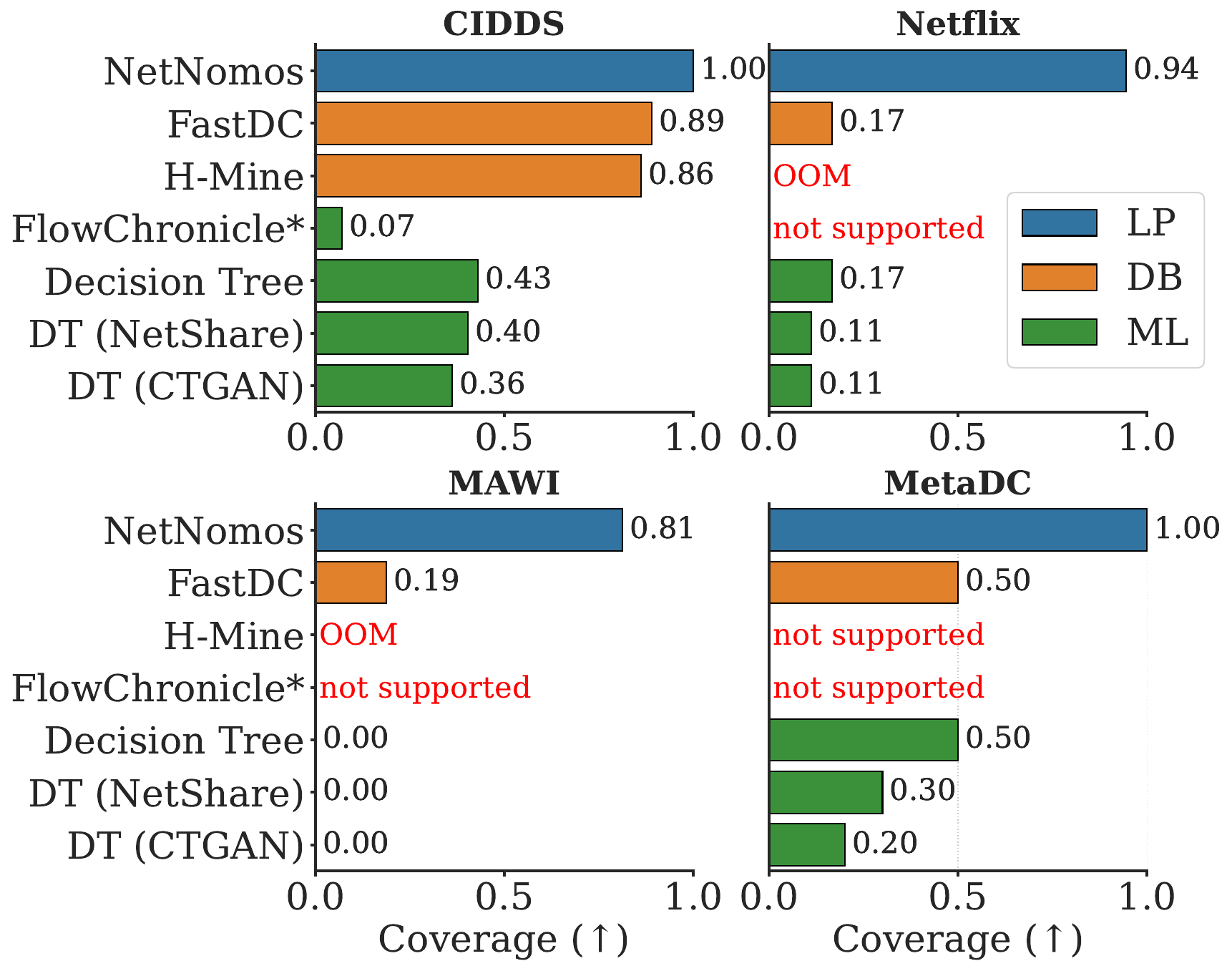}
    \end{adjustbox}
\caption{Expressiveness evaluated under four datasets. \sys is much more expressive than other existing works.
} 
\label{fig:expressiveness2}
\end{figure}

Fig.~\ref{fig:expressiveness2} shows the full results of \sys's expressiveness evaluated on four network datasets.

\section{Scalability of Rule Learning}
\label{apdx:fig-scale}

\begin{figure}[t]
    \centering
    \begin{adjustbox}{width=1\linewidth,center=0pt}
    \includegraphics[width=\linewidth]{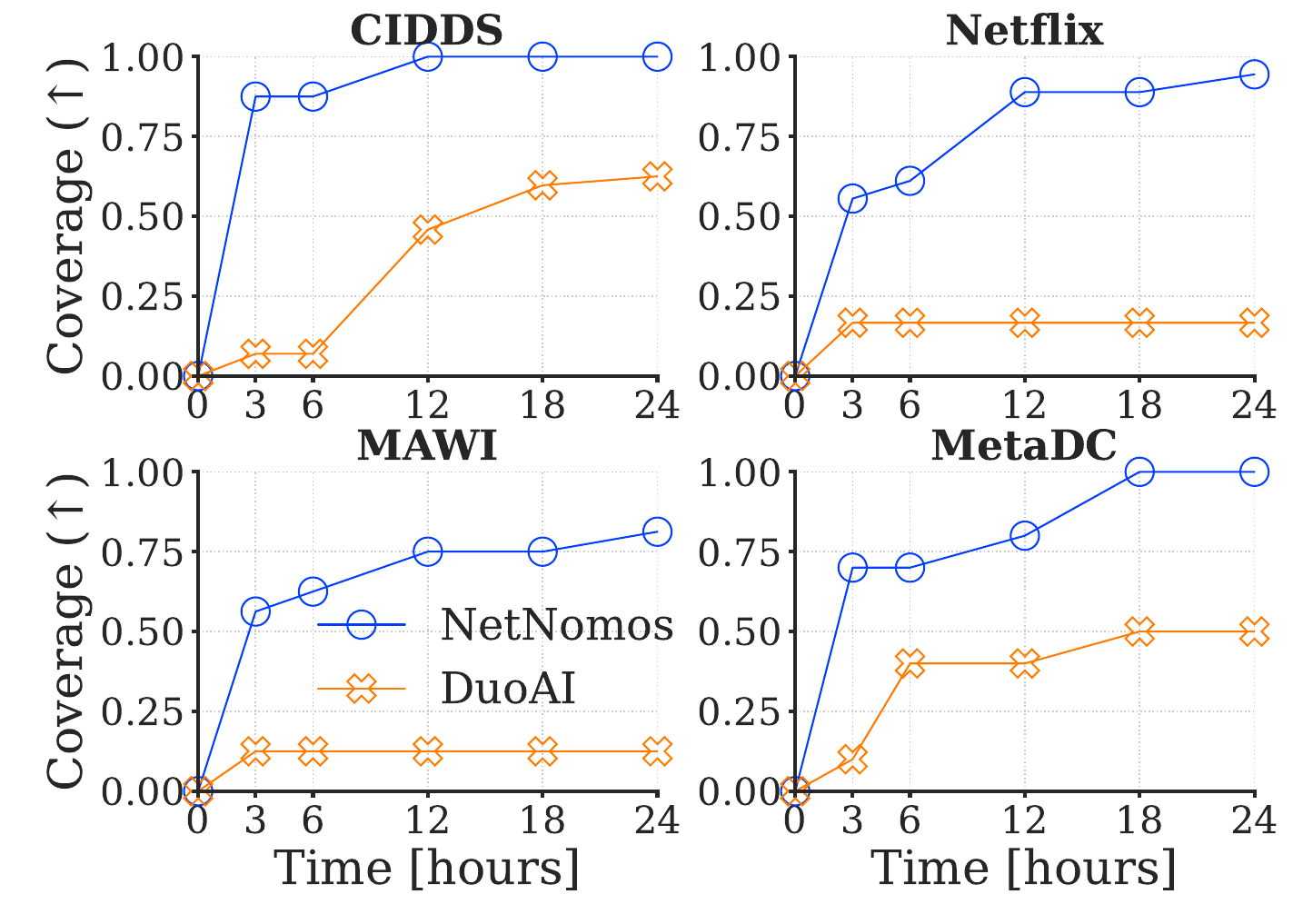}
    \end{adjustbox}
\caption{\sys outscales DuoAI under four evaluated datasets.} 
\label{fig:scalability2}
\end{figure}

Fig.~\ref{fig:scalability2} shows the full results of \sys's learning scalability evaluated on four network datasets.

\section{Rule Filtering Rate}
\label{apdx:filter-rate}

\begin{table}[!h]
\centering
 \begin{adjustbox}{width=0.9\linewidth,center=0pt}    
\begin{tabular}{lrrr}
\toprule
\textbf{Dataset} & \textbf{Learned Rules} & \textbf{Filtered Rules} & \textbf{Filtering Rate}\\
\midrule
CIDDS   & 1,203  & 1,150 & 0.96\\
Netflix & 11,479 & 10,098 & 0.88 \\
MAWI    & 50,989 & 44,859 & 0.88 \\
MetaDC  & 5,934  & 5,322 & 0.91 \\
\bottomrule
\end{tabular}
\end{adjustbox}
\caption{Number of rules learned by \sys after 24h and number of rules filtered out by its semantic filter.
}
\label{tab:selection_rates}
\end{table}

Table~\ref{tab:selection_rates} describes the filtering rate when using a semantic filter against the learned rules associated with the data sets.

\section{LLM Filtering Result}
\label{apdx:llm-filter}

\begin{figure*}[t]
    \centering
    \begin{adjustbox}{width=1\linewidth,center=0pt}
    \includegraphics[width=\linewidth]{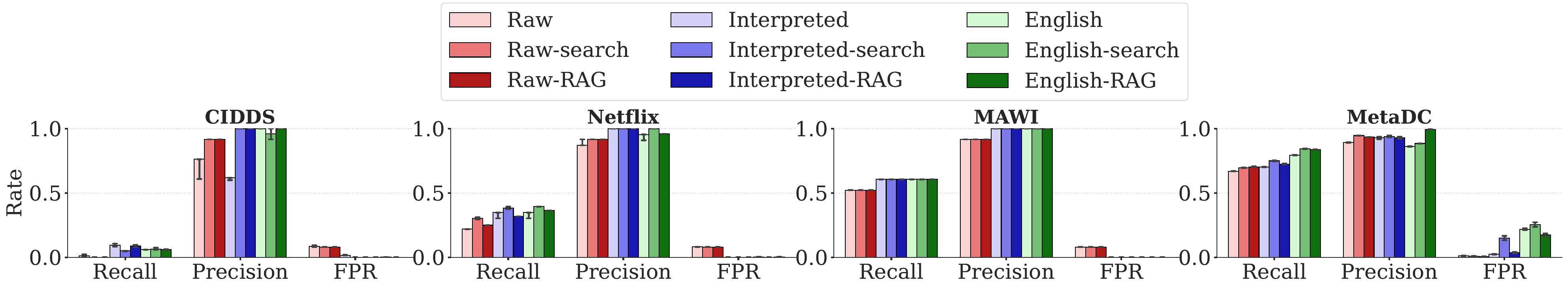}
    \end{adjustbox}
\caption{Similar to the Fig.~\ref{fig:llm_filter}, the use of LLM in \sys is able to filter out meaningless rules when testing under four different datasets.}
\label{fig:llm_filter2}
\end{figure*}

Fig.~\ref{fig:llm_filter2} describes the performance of the LLM filtering under four evaluated datasets. 

\section{Synthetic Data Fidelity}
\begin{figure*}[t]
    \centering
    \begin{adjustbox}{width=0.9\linewidth,center=0pt}
    \includegraphics[width=\linewidth]{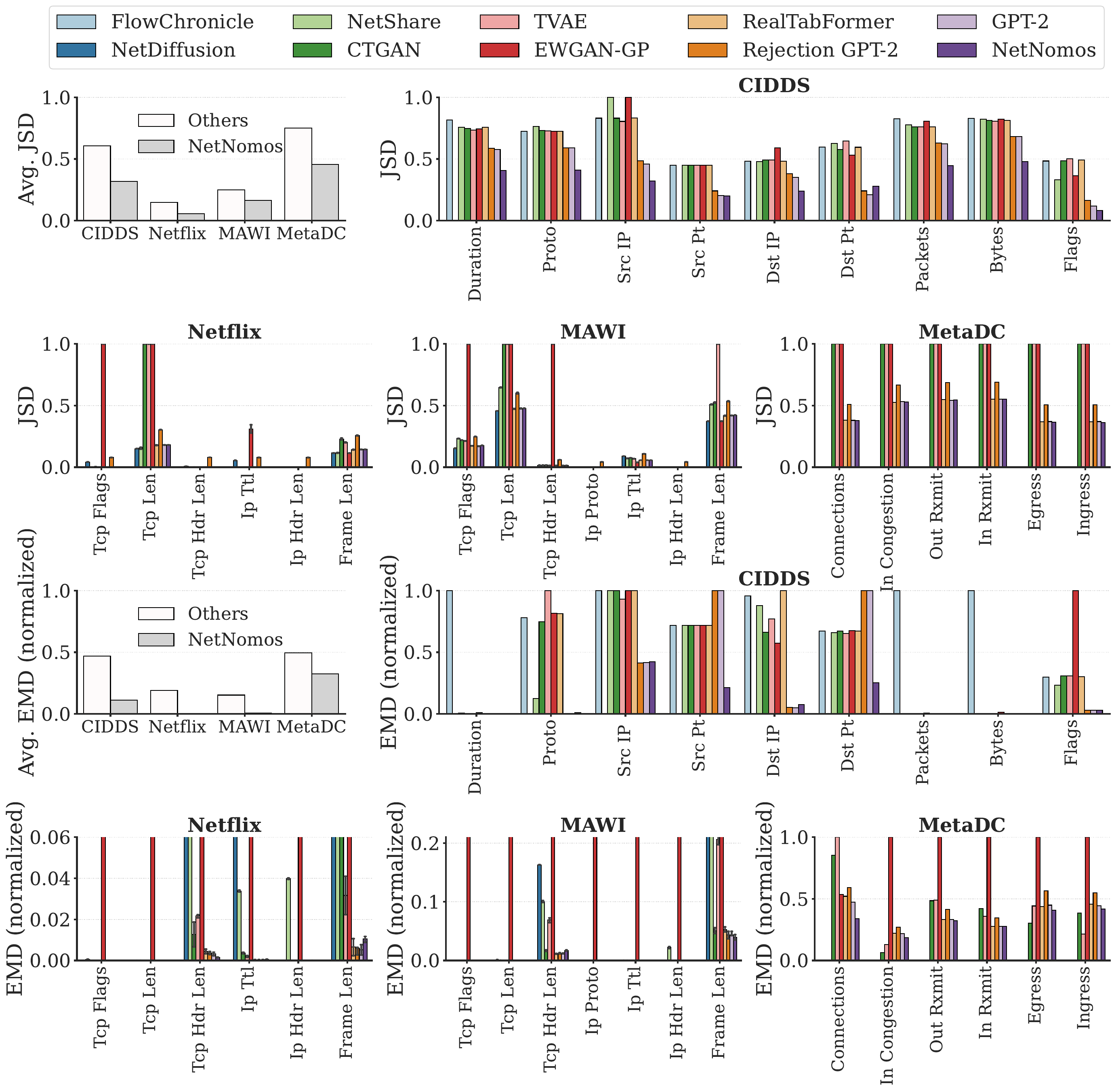}
    \end{adjustbox}
\caption{By imposing learned network rules, \sys generates high-fidelity synthetic data across diverse datasets compared to SOTA data generators.}
\label{fig:syngen2}
\end{figure*}

Fig.~\ref{fig:syngen2} shows the performance of various data generators across datasets.
\label{apdx:syngen}

\section{Traffic Forecasting Performance}
\label{apdx:forecast}

\begin{figure*}[t] 
    \centering
    \begin{adjustbox}{width=0.9\linewidth,center=0pt}
    \includegraphics[width=\linewidth]{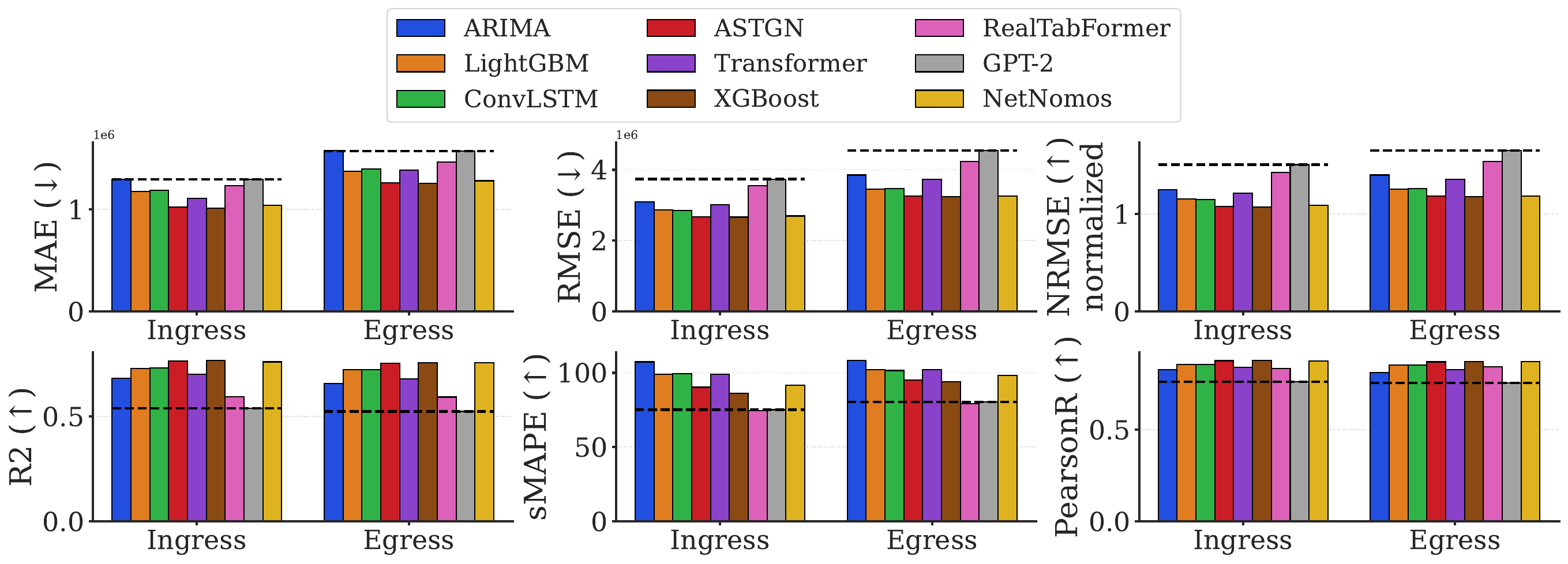}
    \end{adjustbox}
\caption{By enforcing learned rules, \sys improves generic GPT-2 (marked by dashed line) on traffic forecasting by 15.04\%--42.67\% across six metrics and achieve competitive performance against regression and specialized model, ASTGN~\cite{peng2024astgn}.}
\label{fig:forecast2}
\end{figure*}

Fig.~\ref{fig:forecast2} shows traffic forecasting performance evaluated on the MetaDC dataset~\cite{ghabashneh2022millisampler} with six metrics.

\section{Rule Compliance of ML-Generated Data}
\label{apdx:rule-compilance}

\begin{figure*}[t]
    \centering
    \begin{adjustbox}{width=1\linewidth,center=0pt}
    \includegraphics[width=\linewidth]{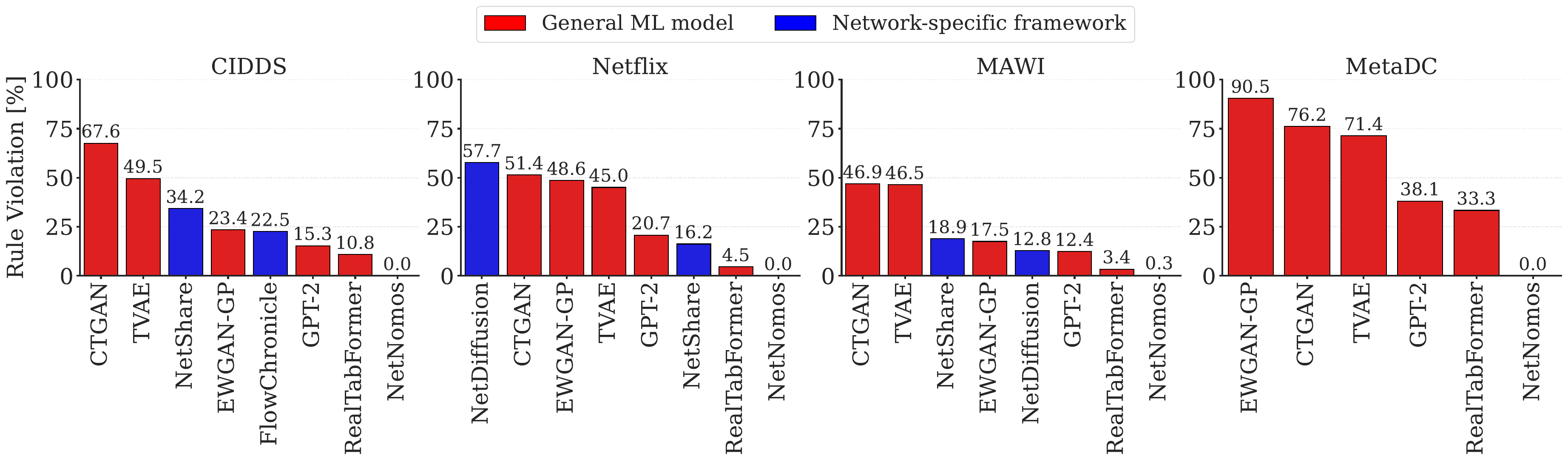}
    \end{adjustbox}
\caption{\sys ensures compliance with the learned rules by enforcing them during inference, achieving zero violations on all datasets except MAWI. In contrast, other frameworks show high violation rates, revealing severe infringements of network semantics in the generated data. Notably, \textit{network-specific frameworks do not display a clear advantage over general models in preserving network semantics, suggesting insufficient integration of domain knowledge.}}
\label{fig:vrank}
\end{figure*}

\begin{figure*}[t]
    \centering
    \begin{adjustbox}{width=0.9\linewidth,center=0pt}
    \includegraphics[width=\linewidth]{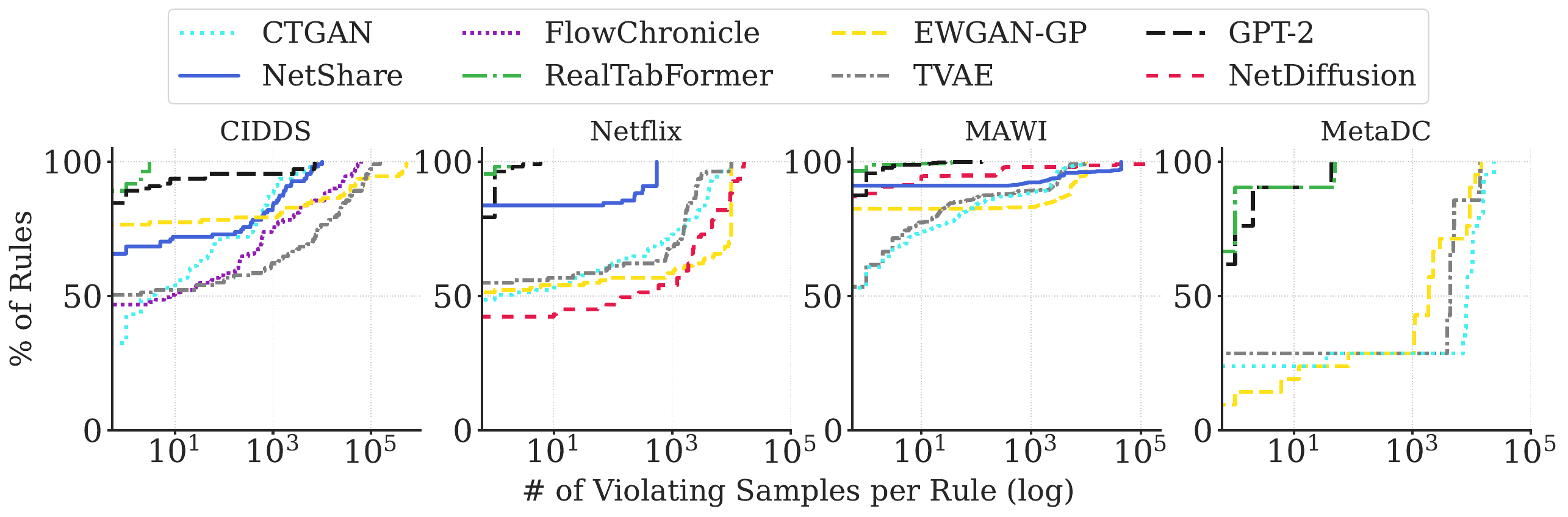}
    \end{adjustbox}
\caption{More than 50\% of the rules are violated by multiple samples from the synthetic data generators. The presence of long tails further indicate limited diversity, as many samples tend to violate the same subset of rules. For instance, on the MetaDC dataset, GPT-2 and RealTabFormer generate considerably more diverse samples than the other models.
}
\label{fig:vstats}
\end{figure*}

Fig.~\ref{fig:vrank} describes rule violation rates of various data generators.
Fig.~\ref{fig:vstats} is the CDF of the number of violating samples per rule.

\section{Prompt Used for Semantic Filtering} \label{apdx:prompt}

We attach the prompt template used for semantic filtering below.

\begin{promptbox}
You are given a list of logical rules extracted from network data.

These rules aim to describe relationships between fields in network data.

\#\#Task

Identify which rules are semantically meaningful. A rule is meaningful only if it reflects real network behavior or expresses a sound integrity constraint.

\#\#Sources of rules

1. Protocol specifications (e.g., RFCs) describing standard behaviors such as port assignments and TCP handshakes.

2. Deployment patterns specific to certain datasets such as no occurrence of public-to-public IP flows.

3. Principles such as queueing, congestion, bandwidth limits, etc.

\#\#Output format

Return a Python-style list of the rule IDs that are consistent. Example:
[0, 2, 5]

Do not return anything else and be critical.

\#\#Rules:
\{rules\}
\end{promptbox}

\section{Testbed}
\label{apdx:testbed}
For fair evaluation, we conduct all the experiments on a two-socket server with 40 logical CPUs of Intel Xeon E5-2660v3 with a frequency of 2.60GHz and  256GiB of DRAM.
Training and inference of ML models are all conducted on an A100 NVIDIA GPU.

\begin{table*}[t]
  \caption{Sample PCAP constraints from our Netflix and MAWI benchmark rulesets. SameBiFlow($\cdot$) and SameDir($\cdot$) are shorthand notations for flow matching.}
  \label{tab:pcap_bench}
  \centering
  \renewcommand{\arraystretch}{1.2}
  \begin{tabularx}{\textwidth}{
      >{\raggedright\arraybackslash}p{0.36\textwidth}
      X
      >{\raggedright\arraybackslash}p{0.16\textwidth}
      >{\raggedright\arraybackslash}p{0.08\textwidth}}
    \toprule
    \textbf{Network Constraint} & \textbf{Meaning} & \textbf{Expressible by} & \textbf{Reference} \\
    \midrule
\begin{minipage}[t]{\linewidth}\vspace{0pt}
  \begin{enumerate}[leftmargin=*, label=\arabic*.]
    \setcounter{enumi}{0}
    \item $\forall t,p: \text{SameBiFlow}(p_t, p_{t+1}, p_{t+2}) \land \textsl{SYN} \in p_t.\texttt{Flags} \land \textsl{SYN-ACK} \in p_{t+1}.\texttt{Flags} \land \textsl{ACK} \in p_{t+2}.\texttt{Flags} \implies p_{t+2}.\texttt{Seq} = p_{t+1}.\texttt{Ack} \land p_{t+1}.\texttt{Ack} = (p_{t}.\texttt{Seq}+1) \land p_{t+2}.\texttt{Ack} = (p_{t+1}.\texttt{Seq}+1)$
  \end{enumerate}
\end{minipage}
&
\textbf{[Protocol]}
  TCP three-way handshake.
&
\sys
&
\cite{jiang2024netdiffusion}, \cite{ma2017netflix}, RFC 9293~\cite{rfc9293}, 1122~\cite{rfc1122}
\\
\midrule

\begin{minipage}[t]{\linewidth}\vspace{0pt}
  \begin{enumerate}[leftmargin=*, label=\arabic*.]
    \setcounter{enumi}{1}
    \item $\forall t,p: p.\texttt{IpVersion} = 4 \lor p.\texttt{IpVersion} = 6$
  \end{enumerate}
\end{minipage}
&
\textbf{[Protocol]}
  Correct IP version number.
&
\sys, H-Mine~\cite{pei2007hmine}, FastDC~\cite{chu2013fastdc,pena2022fastdc++},
Decision Trees
&
IANA~\cite{iana-protocol-numbers}, RFC 791~\cite{rfc791}, 4443~\cite{rfc4443}, 8200~\cite{rfc8200}, 9293~\cite{rfc9293}
\\
\midrule

\begin{minipage}[t]{\linewidth}\vspace{0pt}
  \begin{enumerate}[leftmargin=*, label=\arabic*.]
    \setcounter{enumi}{2}
    \item $\forall t,p: \text{SameBiFlow}(p_t, p_{t+1}) \land \textsl{PSH-ACK} \in p_{t}.\texttt{Flags} \implies \textsl{ACK} \in p_{t+1}.\texttt{Flags} \lor \textsl{RST} \in p_{t+1}.\texttt{Flags}$
  \end{enumerate}
\end{minipage}
&
\textbf{[Protocol]}
  PSH data packet followed by ACK.
&
\sys
&
RFC 1122~\cite{rfc1122}, 9293~\cite{rfc9293}
\\
\midrule

\begin{minipage}[t]{\linewidth}\vspace{0pt}
  \begin{enumerate}[leftmargin=*, label=\arabic*.]
    \setcounter{enumi}{3}
    \item $\forall t,p: \text{SameBiFlow}(p_t, p_{t+1}, p_{t+2}) \land \textsl{SYN} \in p_t.\texttt{Flags} \land \textsl{SYN-ACK} \implies  p_t.\texttt{TcpWinSize} > 0 \land p_{t+2}.\texttt{TcpWinSize} > 0$
  \end{enumerate}
\end{minipage}
&
\textbf{[Protocol]}
  TCP rwnd negotiation. 
&
\sys
&
\cite{ma2017netflix}, RFC 1122~\cite{rfc1122}, 9293~\cite{rfc9293}, 7323~\cite{rfc7323}
\\
\midrule

\begin{minipage}[t]{\linewidth}\vspace{0pt}
  \begin{enumerate}[leftmargin=*, label=\arabic*.]
    \setcounter{enumi}{4}
    \item $\forall t,p: p.\texttt{TcpUrgPointer}>0 \Longleftrightarrow \textsl{URG} \in p.\texttt{Flags} \lor \textsl{RST} \in p.\texttt{Flags}$ 
  \end{enumerate}
\end{minipage}
&
\textbf{[Protocol]}
  Urgent pointer is set if only if URG flag bit is active. 
&
\sys,
FastDC~\cite{chu2013fastdc,pena2022fastdc++},
Decision Trees
&
RFC 9293~\cite{rfc9293}
\\
\midrule

\begin{minipage}[t]{\linewidth}\vspace{0pt}
  \begin{enumerate}[leftmargin=*, label=\arabic*.]
    \setcounter{enumi}{5}
    \item $\forall t,p: p.\texttt{IpHdrLen}\%4 = 0 \land p.\texttt{TcpHdrLen}\%4 = 0$ 
  \end{enumerate}
\end{minipage}
&
\textbf{[Protocol]}
  Header length alignment.
&
\sys
&
\cite{mawi2006traffic}, RFC 791~\cite{rfc791}, 9293~\cite{rfc9293}, 1122~\cite{rfc1122}, 8200~\cite{rfc8200}, 9673~\cite{rfc9673}
\\
\midrule

\begin{minipage}[t]{\linewidth}\vspace{0pt}
  \begin{enumerate}[leftmargin=*, label=\arabic*.]
    \setcounter{enumi}{6}
    \item $\forall t,p: \text{SameDir}(p_t, p_{t+1}) \land \textsl{SYN} \cancel{\in} \{p_t.\texttt{Flags}\cup p_{t+1}.\texttt{Flags}\} \land \textsl{FIN} \cancel{\in} \{p_t.\texttt{Flags}\cup p_{t+1}.\texttt{Flags}\} \implies p_{t+1}.\texttt{Seq} = p_t.\texttt{TcpLen}+p_t.\texttt{Seq}$
  \end{enumerate}
\end{minipage}
&
\textbf{[Protocol]}
  TCP sequence number continuity.  
&
\sys
&
\cite{chu2025netssm}, \cite{jiang2024netdiffusion}, RFC 1122~\cite{rfc1122}, 9293~\cite{rfc9293}
\\
\midrule

\begin{minipage}[t]{\linewidth}\vspace{0pt}
  \begin{enumerate}[leftmargin=*, label=\arabic*.]
    \setcounter{enumi}{7}
    \item $\forall t,p: 0 \leq p.\texttt{IpTtl} \leq 255$
  \end{enumerate}
\end{minipage}
&
\textbf{[Protocol]}
  Valid TTL number. 
&
\sys
&
RFC 791~\cite{rfc791}, 1122~\cite{rfc1122}, 8200~\cite{rfc8200}
\\

    \bottomrule
  \end{tabularx}
\end{table*}
\begin{table*}[t]
  \caption{Sample NetFlow constraints from our CIDDS benchmark ruleset.}
  \label{tab:cidds_bench}
  \centering
  \renewcommand{\arraystretch}{1.2}
  \begin{tabularx}{\textwidth}{
      >{\raggedright\arraybackslash}p{0.36\textwidth}
      X
      >{\raggedright\arraybackslash}p{0.16\textwidth}
      >{\raggedright\arraybackslash}p{0.08\textwidth}}
    \toprule
    \textbf{Network Constraint} & \textbf{Meaning} & \textbf{Expressible by} & \textbf{Reference} \\
    \midrule
\begin{minipage}[t]{\linewidth}\vspace{0pt}
  \begin{enumerate}[leftmargin=*, label=\arabic*.]
    \setcounter{enumi}{0}
    \item $\forall e: e.\texttt{SrcIp} \cancel{\in} \{\textsl{Multicast} \cup \textsl{Broadcast}\} \land e.\texttt{DstIp} \neq \texttt{0.0.0.0}$
  \end{enumerate}
\end{minipage}
&
\textbf{[Protocol]}
  Multicast or broadcast IPs can only appear as destination IP addresses, and wildcard mask can only be used in source IP.
&
\sys
&
RFC 1122~\cite{rfc1122}, 791~\cite{rfc791}, 4443~\cite{rfc4443}, 8200~\cite{rfc8200}
\\
\midrule
\begin{minipage}[t]{\linewidth}\vspace{0pt}
  \begin{enumerate}[leftmargin=*, label=\arabic*.]
    \setcounter{enumi}{1}
    \item $\forall e: e.\texttt{DstPt}  \in \{80, 443\}  \implies \texttt{SrcIp} \in \textsl{Private}$
  \end{enumerate}
\end{minipage}
&
\textbf{[Deployment]}
  Web traffic only originates from within internal network.
&
\sys, H-Mine~\cite{pei2007hmine}, FastDC~\cite{chu2013fastdc,pena2022fastdc++},
Decision Trees
&
\cite{ring2019rulecidds,schoen2024rulecidds,jacobs2022trustee}
\\
\midrule

\begin{minipage}[t]{\linewidth}\vspace{0pt}
  \begin{enumerate}[leftmargin=*, label=\arabic*.]
    \setcounter{enumi}{2}
    \item $\forall e: e.\texttt{DstIp} \in \text{DNS} \implies e.\texttt{DstPt} = 53$
  \end{enumerate}
\end{minipage}
&
\textbf{[Protocol]}
    DNS service employs port 53 by IANA convention.
&
\sys, H-Mine~\cite{pei2007hmine}, FastDC~\cite{chu2013fastdc,pena2022fastdc++},
Decision Trees,
FlowChronicle~\cite{cuppers2024flowchronicle}
&
RFC 768~\cite{rfc768}, 7605~\cite{rfc7605}, IANA~\cite{iana-service-names}
\\
\midrule

\begin{minipage}[t]{\linewidth}\vspace{0pt}
  \begin{enumerate}[leftmargin=*, label=\arabic*.]
    \setcounter{enumi}{3}
    \item $\forall e: e.\texttt{Bytes} \leq 65535\times e.\texttt{Packets}$
  \end{enumerate}
\end{minipage}
&
\textbf{[Protocol]}
    Total amount of transmitted data in a flow is bounded by the number of packets and MTU.
&
\sys
&
RFC 1122~\cite{rfc1122}, 791~\cite{rfc791}, 9293~\cite{rfc9293}, 4443~\cite{rfc4443}
\\
\midrule

\begin{minipage}[t]{\linewidth}\vspace{0pt}
  \begin{enumerate}[leftmargin=*, label=\arabic*.]
    \setcounter{enumi}{4}
    \item $\forall e: e.\texttt{DstPt} \in \{137, 138\} \implies e.\texttt{SrcIp}\in\textsl{Private} \land e.\texttt{DstIp}\in\textsl{Broadcast}$
  \end{enumerate}
\end{minipage}
&
\textbf{[Deployment]}
    NetBIOS flows contain only broadcast packets and use ports 137/138.
&
\sys, H-Mine~\cite{pei2007hmine}, FastDC~\cite{chu2013fastdc,pena2022fastdc++},
Decision Trees
&
\cite{ring2019rulecidds,schoen2024rulecidds}
\\
\midrule

\begin{minipage}[t]{\linewidth}\vspace{0pt}
  \begin{enumerate}[leftmargin=*, label=\arabic*.]
    \setcounter{enumi}{5}
    \item $\forall e: e.\texttt{Proto}\neq \text{TCP} \implies e.\texttt{Flags}=\emptyset$
  \end{enumerate}
\end{minipage}
&
\textbf{[Protocol]}
    Non-TCP flow records do not contain flag information.
&
\sys, H-Mine~\cite{pei2007hmine}, FastDC~\cite{chu2013fastdc,pena2022fastdc++},
Decision Trees
&
RFC 9293~\cite{rfc9293}, 768~\cite{rfc768}, 791~\cite{rfc791}
\\
\midrule

\begin{minipage}[t]{\linewidth}\vspace{0pt}
  \begin{enumerate}[leftmargin=*, label=\arabic*.]
    \setcounter{enumi}{6}
    \item $\forall e: e.\texttt{DstIp}\in \textsl{Public} \Longleftrightarrow e.\texttt{SrcIp} \in \textsl{Private}$
  \end{enumerate}
\end{minipage}
&
\textbf{[Deployment]}
    No public-to-public traffic, since the vantage point was located behind the gateway.
&
\sys, H-Mine~\cite{pei2007hmine}, FastDC~\cite{chu2013fastdc,pena2022fastdc++},
Decision Trees
&
\cite{ring2019rulecidds,schoen2024rulecidds,jacobs2022trustee}
\\
\midrule

\begin{minipage}[t]{\linewidth}\vspace{0pt}
  \begin{enumerate}[leftmargin=*, label=\arabic*.]
    \setcounter{enumi}{7}
    \item $\forall e: e.\texttt{Proto} = \textsl{UDP} \implies e.\texttt{Bytes} \geq 8\times e.\texttt{Packets}$
  \end{enumerate}
\end{minipage}
&
\textbf{[Protocol]}
    A UDP flow entry contains at least one packet.
&
\sys
&
RFC 1122~\cite{rfc1122}
\\

    \bottomrule
  \end{tabularx}
\end{table*}
\begin{table*}[t]
  \caption{Sample constraints from our MetaDC benchmark rulesets.}
  \label{tab:meta_bench}
  \centering
  \renewcommand{\arraystretch}{1.2}
  \begin{tabularx}{\textwidth}{
      >{\raggedright\arraybackslash}p{0.36\textwidth}
      X
      >{\raggedright\arraybackslash}p{0.17\textwidth}
      >{\raggedright\arraybackslash}p{0.07\textwidth}}
    \toprule
    \textbf{Network Constraint} & \textbf{Meaning} & \textbf{Expressible by} & \textbf{Reference} \\
    \midrule
\begin{minipage}[t]{\linewidth}\vspace{0pt}
  \begin{enumerate}[leftmargin=*, label=\arabic*.]
    \setcounter{enumi}{0}
    \item $\forall e: e.\texttt{Ingress} \geq e.\texttt{InRxmit} \land e.\texttt{Egress} \geq e.\texttt{OutRxmit}$
  \end{enumerate}
\end{minipage}
&
\textbf{[Principle]}
  Retransmitted traffic is contained in the total traffic volume.
&
\sys
, FastDC~\cite{chu2013fastdc,pena2022fastdc++},
&
--------
\\
\midrule

\begin{minipage}[t]{\linewidth}\vspace{0pt}
  \begin{enumerate}[leftmargin=*, label=\arabic*.]
    \setcounter{enumi}{1}
    \item $\forall t, e: e.\texttt{Congestion}_t > 0 \implies e.\texttt{Ingress}_t > 0$
  \end{enumerate}
\end{minipage}
&
\textbf{[Principle]}
  Congestion markings indicate ingress traffic.
&
\sys,
FastDC~\cite{chu2013fastdc,pena2022fastdc++},
Decision Trees
&
--------
\\
\midrule

\begin{minipage}[t]{\linewidth}\vspace{0pt}
  \begin{enumerate}[leftmargin=*, label=\arabic*.]
    \setcounter{enumi}{2}
    \item $\forall t,e: e.\texttt{Connections}_t > 26700 \implies (e.\texttt{InCongession}_t > 0) \lor (e.\texttt{InRxmit}_t > 0)$  
  \end{enumerate}
\end{minipage}
&
\textbf{[Principle]}
  Large number of active connections (p90) indicate congestion or retransmission (incast).
&
\sys,
Decision Trees
&
--------
\\
\midrule

\begin{minipage}[t]{\linewidth}\vspace{0pt}
  \begin{enumerate}[leftmargin=*, label=\arabic*.]
    \setcounter{enumi}{3}
    \item $\forall t, e: e.\texttt{Connections}_t > 0 \implies (e.\texttt{Ingress}_t > 0) \lor (e.\texttt{Egress}_t > 0)$
  \end{enumerate}
\end{minipage}
&
\textbf{[Principle]}
  Active connection leads to traffic.
&
\sys,
Decision Trees
&
--------
\\
\midrule

\begin{minipage}[t]{\linewidth}\vspace{0pt}
  \begin{enumerate}[leftmargin=*, label=\arabic*.]
    \setcounter{enumi}{4}
    \item $\forall t,e: \sum_{k=1}^{K=50} e.\texttt{InCongestion}_{t+k} > 206305 \implies \exists i \in [1, 5]: e.\texttt{IngressRate10ms}_{t+i} > 5357$
  \end{enumerate}
\end{minipage}
&
\textbf{[Principle]}
  Heavy congestion implies micro-bursts.
&
\sys
&
--------
\\

    \bottomrule
  \end{tabularx}
\end{table*}

\end{document}